\documentclass[11pt]{article}
\pdfoutput=1

\usepackage{amsmath,amsfonts,amssymb,epsfig}
\usepackage{slashed} 
\usepackage[width=0.9\textwidth,margin={30pt,30pt},font=small,labelfont=bf]{caption}
\numberwithin{equation}{section}
 
\usepackage{hyperref}
\usepackage{enumerate} 

\usepackage{bbold}
\usepackage{xcolor}
\hypersetup{
    colorlinks,
    linkcolor={red!0!black},
    citecolor={blue!50!black},
    urlcolor={blue!80!black}
}
\usepackage{array}
\usepackage{subfig}
\usepackage{subcaption}
\usepackage{graphicx}
\usepackage{cancel}
\usepackage{color}
\usepackage{bm}
\usepackage{multirow}
\usepackage{multicol}
\usepackage{xspace}
\usepackage{tcolorbox}
\usepackage{adjustbox}
\usepackage{tikz}

\usepackage{tikz-feynman}
\tikzfeynmanset{compat=1.0.0}

\usepackage{cite}
\usepackage{float}

\newcommand{\exclude}[1]{}
\newcommand{\sigmabar}{\bar{\sigma}}

\newcommand{\al}{\alpha}

\newcommand{\psib}{\bar{\psi}}

\newcommand{\bone}{\boldsymbol{1}}
\newcommand{\btwo}{\boldsymbol{2}}
\newcommand{\bthree}{\boldsymbol{3}}

\newcommand{\bp}{\boldsymbol{p}}

\newcommand{\dotalpha}{\Dot{\alpha}}
\newcommand{\dotbeta}{\Dot{\beta}}

\begin{document}
  
\hypersetup{pageanchor=false}
\begin{titlepage}

\begin{center}

\hfill UMN-TH-4425/25 \\
\hfill FTPI-MINN-25-06\\

\vskip 0.6in

{\LARGE \bfseries Supergravity from the Bottom Up} \\
\vskip .4in

{\normalsize\bf Tony Gherghetta,$^{1,a}$} \let\thefootnote\relax\footnote{$^a$tgher@umn.edu}
{\normalsize\bf Wenqi Ke$^{1,2,b}$}
\footnote{$^b$wke@umn.edu}
\vskip .1in
\begin{tabular}{ll}
$^{1}$ & \!\!\!\!\!\emph{\footnotesize School of Physics and Astronomy, University of Minnesota, 116 Church St SE, }\\
& \!\!\!\!\!\emph{\footnotesize Minneapolis, Minnesota 55455, USA}\\
$^{2}$ & \!\!\!\!\!\emph{\footnotesize  William I. Fine Theoretical Physics Institute, School of Physics and Astronomy,}\\[-.15em]
& \!\!\!\!\!\emph{\footnotesize University of Minnesota, 116 Church St SE, Minneapolis, Minnesota 55455, USA}\\
\end{tabular}

\end{center}
\vskip .4in

\begin{abstract}
\noindent
We employ on-shell methods to construct scattering amplitudes and derive effective theories involving massive spin-3/2 fermions interacting with spin 0, 1 and 2 bosons.  The four-point massive amplitudes are constructed using an all-line-transverse momentum shift, assuming that in the massless limit, three-point interactions are smooth and the Ward identity is satisfied. For a Majorana spin-3/2 fermion with mass $m_{3/2}$, we show that interactions with only spin 0 and massive spin-1 bosons do not lead to an effective theory valid up to a cutoff $\Lambda \gg m_{3/2}$ that is independent of particle masses. Instead, adding an interaction with a spin-2 graviton gives rise to four-point amplitudes with a Planck scale unitarity cutoff that reproduces well-known results from $N=1$ supergravity, such as $F$-term breaking with a complex scalar and $D$-term breaking with an additional massive photon. These bottom-up results are then extended to two Majorana spin-3/2 fermions where an interacting effective theory valid up to $\Lambda \gg m_{3/2}$ again requires the introduction of the spin-2 graviton. Unitarity up to the Planck scale is then achieved when the two Majorana spin-3/2 fermions have unequal masses, and necessarily couple to {\it two} massive spin-1 states  corresponding to the spontaneous breaking of $N=2$ supergravity to $N=0$. Our results, obtained from the bottom-up and without any Lagrangian, imply that broken supergravity is the unique, effective theory involving interactions of massive spin-3/2 fermions valid up to a cutoff $\Lambda \gg m_{3/2}$ that does not depend on particle masses.

\end{abstract}

\end{titlepage} 

\tableofcontents

\newpage

\section{Introduction}

On-shell methods provide a particularly simple and efficient way to determine scattering amplitudes without needing a Lagrangian or the concomitant Feynman rules. By imposing locality and Lorentz symmetry, three-point amplitudes that encode the primitive interactions between on-shell particles, provide the building blocks to generate higher-point amplitudes via recursion relations. This bottom-up construction of interactions provides a way to unveil theoretical structures and symmetries.

Amongst the elementary states allowed by the little-group symmetry, spin-3/2 particles, are yet to be observed in nature, but play a fundamental role in supergravity and string theories. Although their interactions have been well-studied from the top down, primarily with supersymmetric Lagrangians, the analysis is necessarily  complicated due to the inherent redundancy of describing a spin-3/2 state with a vector-spinor field, $\psi_\mu$, where $\mu$ is the Lorentz index. Consequently, the off-shell state $\psi_\mu$ must be accompanied with extra constraint conditions that remove the unphysical degrees of freedom. It is clear that this complexity can be avoided by employing on-shell methods.

The use of on-shell methods to study massless spin-3/2 states was previously considered in Ref.~\cite{McGady:2013sga}, which concluded that massless spin-3/2 states must interact gravitationally with a naturally emergent supersymmetry. This complements a much earlier study in Ref.~\cite{Grisaru:1977kk} that used soft theorems to reach the same conclusion. Instead, our focus will be on {\it massive} spin-3/2 states with mass $m_{3/2}$, where we perform a systematic, bottom-up construction of the allowed spin-3/2 interactions with spin $\leq 2$ states that lead to theories with a unitarity cutoff $\Lambda\gg m_{3/2}$, independent of the particle mass. 
To build up these consistent theories from three-point interactions (containing up to one derivative), we assume that in the massless limit ($m_{3/2}\rightarrow 0$), the three-point interactions are smooth and satisfy the Ward identity. Under these mild assumptions we will recover the interactions of (broken) supergravity, generalizing the previous massless spin-3/2 studies~\cite{Grisaru:1977kk,McGady:2013sga}.

The important technical step that will be used to build up massive four-point amplitudes is an all-line-transverse (ALT) momentum shift, first proposed in Ref.~\cite{Ema:2024vww}. When this is applied to a Majorana spin-3/2 fermion, the four-point contact term is uniquely fixed. By further adding a complex scalar, the leading $E^4$ divergent terms in the amplitude are canceled, leaving $E^2$ terms with a Planck scale unitarity cutoff. This reproduces the Polonyi model with $F$-term breaking in minimal $N=1$ supergravity~\cite{Gherghetta:2024tob}. 

In this work, we generalize the massive Majorana spin-3/2 fermion to interact with a massive spin-1 boson in addition to the spin-0 complex scalar. While it may appear that this could give rise to a consistent theory valid up to a cutoff $\Lambda\gg m_{3/2}$ which does not depend on the particle masses,  
we show that this is not possible! This is because the  amplitude with longitudinal states always grows as  $E^4$ in the high-energy limit. 
Instead, by introducing a spin-2 graviton with gravitational interaction, $\kappa \equiv 2/M_P$, where $M_P$ is the Planck scale, we are able to satisfy the Ward identity and obtain perturbative unitarity up to order $\kappa^2 E^2$ terms, thereby realizing a unitarity cutoff independent of particle masses. The dimensionful, gravitational coupling $\kappa$ plays a crucial role in realizing a smooth, massless limit. In this way, we again recover $N=1$ supergravity but now with an additional $D$-term breaking associated with the massive vector field where both the superHiggs and Higgs mechanisms emerge.

The bottom-up construction can be extended to two Majorana spin-3/2 states or a massive Dirac spin-3/2 state. In this case, the Dirac spin-3/2 state can couple to a massless U(1) gauge boson, except that the Ward identity cannot be satisfied. Again, by introducing a spin-2 graviton, the Ward identity can be restored in the massless limit, provided the U(1) coupling, $e=\frac{m_{3/2}}{\sqrt{2}M_P}$ (assuming the minimal gravitational coupling). This corresponds to softly-broken $N=2$ supergravity
where the theory remains supersymmetric in the high-energy limit but is not necessarily unitary to ${\cal O}(\kappa^2E^2)$. In fact, the leading $E^4$ behavior of the four-point, spin-3/2 amplitude cannot be canceled with scalar couplings, implying that the unitarity cutoff cannot be increased to the Planck scale. This provides an on-shell derivation of the well-known result that in this model with a charged, massive gravitino and a massless graviphoton, supergravity is only softly, but not spontaneously broken \cite{Deser:1977uq,Deser:2001dt}. A spontaneous breaking of extended supergravity in Minkowski spacetime requires a spontaneous breaking of gauged isometries \cite{Andrianopoli:2002rm}, which generates a mass for the gauge bosons.  Therefore, a crucial ingredient for unitarity is to have a massive spin-$1$ boson.

To derive this fact with on-shell methods, our final example then considers two massive Majorana spin-3/2 fermions coupling to a {\it massive} photon. Again the spin-2 graviton is required to satisfy the Ward identity in the massless limit. Interestingly, there are now solutions where unitarity is maintained to ${\cal O}(\kappa^2 E^2)$. Although unitarity can still not be achieved for the Dirac spin-$3/2$ state, we instead find that the Majorana spin-$3/2$ masses must be unequal where the spin-$3/2$ fermions necessarily interact with two spin-1 bosons with vector and axial-vector couplings. One special solution has mass relations that
correspond to $N=2$ supergravity spontaneously broken to $N=0$. It is also possible to partially break $N=2$ to $N=1$ supergravity in Minkowski spacetime~\cite{Ferrara:1995xi}, where one of the spin-$3/2$ fermions remains massless, and the other massive spin-3/2 fermion is part of the same multiplet as the massive photon. In this case, our unitarity solutions for $N=1$ apply, but the massive spin-1 boson is necessary for satisfying the Ward identity, and we show from  
the smooth massless limit of the massive amplitude that the photon has the same mass as the massive spin-$3/2$ fermion. These known supergravity results are much more simply obtained using on-shell methods, avoiding the use of supersymmetric Lagrangians with off-shell fields.

This paper is organized as follows. In Section~\ref{sec2}, we provide a brief review of the framework, including the spinor-helicity formalism as well as the recursion relation, and derive the on-shell three-point amplitudes relevant for this work.  In Section~\ref{sec3}, we construct $N=1$ supergravity from the bottom-up, first starting with spin-0 and spin-1 bosons that couple to a spin-$3/2$ fermion, then adding a spin-$2$ graviton. In Section~\ref{sec4}, we generalize the particle content to include two Majorana spin-$3/2$ fermions that are charged under a $U(1)$ symmetry. By the recursion relation, we recover the couplings and contact terms of minimal $N=2$ supergravity. We then comment on the high-energy behavior of the amplitude. In Section~\ref{sec5}, we construct the interactions of two Majorana spin-3/2 fermions with a massive photon and then include a graviton. We also provide one special solution  that is  unitarity up to ${\cal O}(\kappa^2 E^2)$. Finally, we complement this work with appendices. In Appendix \ref{app1}, we provide more details about the on-shell formalism, the conventions and useful relations. Appendix \ref{app:shift} summarizes the ALT shift that is employed in the main text.  In Appendix \ref{app:contact}, we present further details of the calculations to obtain the spin-3/2 contact terms from photon exchange. Finally, in Appendix~\ref{app:sugra}, we write the minimal $N=2$ supergravity Lagrangian for reference.

\section{The on-shell framework}\label{sec2}
\subsection{Spinor-helicity   and recursion relation}
The massive spinor-helicity formalism \cite{Arkani-Hamed:2017jhn} generalizes the massless formalism with particles transforming in the little group $SU(2)$. For a massive particle with momentum    $p_{\alpha\dotalpha}$, we can decompose 
$ {p}_{\alpha\dotalpha}\equiv\left| \boldsymbol{p}^I\right>\left[ \boldsymbol{p}_I\right|\,, $
where $I=1,2$ are the $SU(2)$ little-group indices. In the following, we drop the $I$ indices when no ambiguity arises.

It is convenient to choose a special $SU(2)$ basis where the momentum can be written in terms of the two-component spinors $\lambda,\tilde{\lambda},\eta ,\tilde{\eta}$. This can be seen in \eqref{decomp-lam} of Appendix~\ref{app1} where we give a more detailed review of the formulas in Appendix \ref{app:spinor}, and explicit kinematics  in Appendix \ref{app:explicitkin}. In the high-energy limit $E\gg m$, where $m$ is the particle mass, these spinors behave as  $\lambda,\Tilde{\lambda}\sim \sqrt{E}$, $\eta,\Tilde{\eta}\sim \frac{m}{\sqrt{E}}$.

The spinors $\lambda,\eta,\tilde{\lambda},\tilde{\eta}$ can be used to represent different helicities of a particle, because $\lambda,\tilde{\eta}$ have helicity weight $-1/2$ and  $\tilde{\lambda}, {\eta}$ have helicity $+1/2$. The polarization vector for a massive spin-1 boson is given by
\begin{equation}
\varepsilon_\mu^{IJ}=\frac{1}{\sqrt{2}m}
\left<\bp\right|\sigma_\mu\left|\bp\right],\qquad   \varepsilon^{IJ}_{\alpha\Dot{\alpha}}=
\frac{\sqrt{2}}{m}
\left| \bp\right>_\alpha\left[\bp\right|_{\Dot{\alpha}}\,,
\label{eq:massivepolvect}
\end{equation}
where $IJ$ are symmetrized. In the basis \eqref{decomp-lam}, the helicities $(\pm,0)$ are then various combinations of $\lambda,\eta,\tilde{\lambda},\tilde{\eta}$, namely
\begin{equation}
    \varepsilon^+_{\alpha\dotalpha}= \frac{\sqrt{2}}{m} \eta _\alpha\Tilde{\lambda} _{\dotalpha}
  ,\quad \varepsilon^-_{\alpha\dotalpha}= \frac{\sqrt{2}}{m} \lambda _\alpha \Tilde{\eta} _{\dotalpha} ,\quad  
\varepsilon_{\alpha\dotalpha}^0=\frac{1}{m}
(\lambda_\alpha\Tilde{\lambda}_{\dotalpha}+\eta_\alpha\Tilde{\eta}_{\dotalpha})\,.
\label{eq:masslesspolvect}
\end{equation}
Likewise, one can obtain the polarization vector of any spin and helicity. 

In the massless case, using little-group covariance, the three-point amplitudes are uniquely determined by the external helicities up to a global constant (see Eq~\eqref{masslesscplus}-\eqref{masslesscminus}). Similar arguments can be applied to the massive case, and a systematic classification of three-point amplitudes for any spin and mass was derived in \cite{Arkani-Hamed:2017jhn}. In Section~\ref{threepoints}, we will apply this formalism to derive the spin-$3/2$ three-point amplitudes, which will then be used to compute the four-point amplitudes in the following sections. This construction is based on the recursion relation, that is reviewed in Appendix~\ref{app:shift}. The recursion relation relies on a complex momentum shift $p_i\rightarrow \hat{p}_i= p_i+ z r_i$, and the $n$-point amplitude also becomes a function of $z$, ${\cal A}_n\rightarrow \widehat{\cal A}_n(z)$. Using the Cauchy theorem, locality, and pole properties, one can write the amplitude in a factorized form \eqref{npointconstruction}. 

For the amplitude to be \textit{constructible}, it is crucial that $\widehat{\cal A}(z)$ vanishes at complex infinity, and therefore no boundary term $B_\infty$ arises. In that case, the four-point amplitude can be obtained by three-point subamplitudes given in \eqref{four-point-fact}.
In Ref.~\cite{Ema:2024vww}, a massive momentum shift (ALT) is introduced which has been shown to generate no boundary term for  the amplitudes in QED \cite{Ema:2024vww}, the electroweak theory \cite{Ema:2024rss} and the two-derivative $N=1$ supergravity \cite{Gherghetta:2024tob}. The ALT shift and the arguments for constructibility are reviewed in Appendix~\ref{app:shift}.

\subsection{Spin 3/2 interacting with spin 0, 1, 2}\label{threepoints}

Since our goal is to reconstruct gravitino amplitudes in supergravity from the bottom up, 
we derive the on-shell three-point amplitudes, following \cite{Arkani-Hamed:2017jhn}, involving two massive spin-$3/2$ fermions with bosons of spin $\leq 2$. These three-point amplitudes will then  be constrained by Fermi statistics. The particles considered in this work are summarized below: 
\begin{center}
\begin{tabular}{|c|c|c|}
\hline
& spin & mass\\
\hline\hline
scalar, $S$ & 0 &$m_S$\\
pseudoscalar, $P$ & 0 &$m_P$\\
gauge boson, $A_\mu (B_\mu)$ & 1 & $0\,(m_B)$\\ 
\hline\hline
gravitino, $\psi_\mu\, (\chi_\mu)$ & $\frac{3}{2}$ & $m_{\psi}\, (m_\chi)$\\
graviton, $h_{\mu\nu}$& 2& $0$\\
\hline
\end{tabular}  
\label{tableparticle}
\end{center}
The gravitinos $\psi_\mu$ and  $\chi_\mu$ are Majorana fermions. In the spinor notation, we identify the two spin-3/2 states by $\boldsymbol{2}$, $\boldsymbol{3}$ and the boson by $\boldsymbol{1}$. For the spin-0 and 2 interactions, $\boldsymbol{2}$, $\boldsymbol{3}$ are identical Majorana fermions ($\psi_\mu$), while for spin-1 boson the interaction with the two spin-3/2 states can be either with Majorana ($\psi_\mu$) or Dirac ($\psi_\mu,\chi_\mu$) fermions depending on the spin-$1$ mass.  

Using the classification in \cite{Arkani-Hamed:2017jhn}, the three-point amplitudes are obtained for a given external particle mass and spin. The effective couplings of a Majorana spin-$3/2$ to spin $0$ and $2$ were derived in \cite{Gherghetta:2024tob} and we start by briefly reviewing these results. 

When all external particles are massive, the three-point amplitudes are combinations of $3$ $|\boldsymbol{2})$'s, $3$ $|\boldsymbol{3})$'s, and $2s$ $|\boldsymbol{1})$'s for a spin $s$ boson. The notation $|\boldsymbol{i})$ means either a square or angle bracket. 
When $s=0$, this simply results in four spinor structures:
\begin{equation}
    \left[\btwo\bthree\right]^3,\quad \left<\btwo \bthree\right>^3,\quad\left<\btwo\bthree\right>^2\left[\btwo\bthree\right],\quad\left<\btwo\bthree\right>\left[\btwo\bthree\right]^2\,.
    \label{basis-phipsipsi}
\end{equation}
All of them are antisymmetric under $\btwo\leftrightarrow \bthree$, which is consistent with Fermi statistics. 
Furthermore, parity acts as $    \left[\btwo\bthree\right]^3 \leftrightarrow  \left<\btwo \bthree\right>^3$, $  \left<\btwo\bthree\right>^2\left[\btwo\bthree\right] \leftrightarrow\left<\btwo\bthree\right>\left[\btwo\bthree\right]^2$, so different parity-even or odd combinations will be translated to either scalar or pseudoscalar couplings.

If one of the particles is massless and the other two have equal and non-vanishing mass, the stripped amplitude is given by
\begin{equation}
    M^h_{\{\alpha_1\alpha
    _2\alpha_3\},\{\beta_1\beta_2\beta_3\}}=\sum_{i=0}^3g_ix^h\left[\lambda_1^i\left(\frac{p_2\tilde{\lambda}_1}{m}\right)^i\varepsilon^{3-i}\right]_{\{\alpha_1\alpha
    _2\alpha_3\},\{\beta_1\beta_2\beta_3\}}\,,
    \label{nima-3}
\end{equation}
where $g_i$ are coupling constants, and $h$ is the helicity of the massless particle. The three-point amplitude   is given by the contraction of \eqref{nima-3} with the spin-$3/2$ spinors. The ``$x$-factor'' is defined by
\begin{equation}
    x\lambda_{1\alpha}=\frac{p_{2\al\dotalpha}}{m}\Tilde{\lambda}_1^{\dotalpha},\qquad \frac{\tilde{\lambda}_{1}^{\dotalpha}}{x}=\frac{p_2^{\dotalpha\alpha}}{m}
    \lambda_{1\alpha}\,.
    \label{xfactordef}
\end{equation} 
 
When $s=2$, corresponding to a massless graviton, we can use \eqref{nima-3} and the identities \eqref{xrelation}. For a helicity $\pm2$ graviton, the three-point amplitude becomes 
\begin{equation}
\begin{aligned}
{\cal A}_h^{(+2)}=&\left(\left<\btwo\varepsilon^+_1\bthree\right]+\left<\bthree\varepsilon^+_1\btwo\right]\right)^2\left({ g}_0^{(+2)} \left<\btwo\bthree\right> +{ g}_1^{(+2)} [\btwo\bthree]\right) 
 + { g}_2^{(+2)} \left<\btwo\bthree\right>[1\btwo]^2 [1\bthree]^2\\&+{ g}_3^{(+2)} [\btwo\bthree ][1\btwo]^2 [1\bthree]^2\,,
\end{aligned}
\label{gravitonplus}
\end{equation}
\begin{equation}
    \begin{aligned}
        {\cal A}_h^{(-2)} =&\left(\left<\btwo\varepsilon^-_1\bthree\right]+\left<\bthree\varepsilon^-_1\btwo\right]\right)^2\left({ g}_0^{(-2)} [\btwo\bthree]+{ g}_1^{(-2)} \left<\btwo\bthree\right>\right) + { g}_2^{(-2)} \left[\btwo\bthree\right]\left<1\btwo\right>^2 \left<1\bthree\right>^2\\
        &+ { g}_3^{(-2)} \left<\btwo\bthree \right>\left<1\btwo\right>^2 \left<1\bthree\right>^2\,.
    \end{aligned} 
    \label{psipsihminus}
    \end{equation} 
In particular, the ${ g}_0^{(\pm2)}$ amplitude originates from the first term in the expansion \eqref{nima-3}, which is often called the \textit{minimal coupling}. Note that Eqs.~\eqref{gravitonplus}-\eqref{psipsihminus} are also antisymmetric under $\btwo\leftrightarrow \bthree$. 

When $s=1$, Fermi statistics impose non-trivial constraints on the on-shell amplitudes. In the case of a massless spin-$1$ boson,  after expanding \eqref{nima-3}, the amplitude is given by
\begin{equation}
\begin{aligned}{\cal A}_A^{(+1)}
=&\left(\left<\btwo\varepsilon^+_1\bthree\right]+\left<\bthree\varepsilon^+_1\btwo\right]\right) \left[g_0^{(+1)} \left<\btwo\bthree\right>^2  +g_1^{(+1)} \left<\btwo\bthree\right>\left[\btwo\bthree\right]+g_2^{(+1)} \left[\btwo\bthree\right]^2\right]\\&+g_3^{(+1)}   \left[1\btwo \right] \left[1\bthree \right]\left( \left[\btwo\bthree\right]- \left<\btwo\bthree\right>\right)^2\,, 
\end{aligned}\label{gravphoton}
\end{equation} 
for a positive ($+1$) helicity photon and where the amplitude with a negative helicity ($-1$) photon is obtained by exchanging angle and square brackets, as well as $\varepsilon_1^+\rightarrow\varepsilon^-_1$. However, Eq.~\eqref{gravphoton} is now symmetric under $\btwo\leftrightarrow \bthree $ which should not be the case with identical Majorana fermions. Therefore, massive Majorana spin-3/2 fermions cannot couple to a massless spin-1 boson. Instead, a nonzero coupling is only possible 
if the spin-3/2 fermions are Dirac, in other words, $\btwo$ and $\bthree$ are no longer identical, so antisymmetry is not required. 

On the other hand, when the spin-1 is massive, the three-point amplitude is a  product of $\left(\bone\btwo\right)$, $\left(\bone\bthree\right)$, $\left(\btwo\bthree\right)$, $\left(\btwo\bthree\right)$. This yields the following twelve structures
\begin{equation}
\begin{aligned}
 & \left(\left<\bone\btwo\right>,\left[\bone\btwo\right]\right)\otimes\left(\left<\bone\bthree\right>,\left[\bone\bthree\right]\right)\otimes\left(\left<\btwo\bthree\right>^2,\left[\btwo\bthree\right]^2,\left<\btwo\bthree\right>\left[\btwo\bthree\right]\right)\,.
    \end{aligned}
    \label{generalmassive1}
\end{equation}
They are not all independent. In fact, two of them can be expressed in terms of the other amplitudes by applying the Schouten identities (see \eqref{schouten}). This gives
\begin{equation}
\begin{aligned}
& \left<\btwo\bthree\right> \left\{m_1 \left<\bone\btwo\right> \left<\bone\bthree\right>\left[\btwo\bthree\right]+m_2 \left<\bone\btwo\right> \left[\bone\bthree\right]\left<\btwo\bthree\right>+m_3 \left[\bone\btwo\right]\left<\bone\bthree\right>\left<\btwo\bthree\right>\right\}
   \\&\qquad= \left<\btwo\bthree\right> \left\{ m_1 \left[\bone\btwo\right] \left[\bone\bthree\right]\left<\btwo\bthree\right>+m_2 \left[\bone\btwo\right] \left<\bone\bthree\right>\left[\btwo\bthree\right]+m_3 \left<\bone\btwo\right>\left[\bone\bthree\right]\left[\btwo\bthree\right]\right\}\,,
   \end{aligned}\label{schouten1}
\end{equation}
which eliminates 
$\left[\bone\btwo\right]\left[\bone\bthree\right]\left<\btwo\bthree\right>^2$, and 
 \begin{equation}\begin{aligned}
& \left[\btwo\bthree\right] \left\{m_1 \left<\bone\btwo\right> \left<\bone\bthree\right>\left[\btwo\bthree\right]+m_2 \left<\bone\btwo\right> \left[\bone\bthree\right]\left<\btwo\bthree\right>+m_3 \left[\bone\btwo\right]\left<\bone\bthree\right>\left<\btwo\bthree\right>\right\}
   \\&\qquad= \left[\btwo\bthree\right] \left\{ m_1 \left[\bone\btwo\right] \left[\bone\bthree\right]\left<\btwo\bthree\right>+m_2 \left[\bone\btwo\right] \left<\bone\bthree\right>\left[\btwo\bthree\right]+m_3 \left<\bone\btwo\right>\left[\bone\bthree\right]\left[\btwo\bthree\right]\right\}\,,
   \end{aligned}\label{schouten2}
    \end{equation}
which eliminates 
$\left<\bone\btwo\right>\left<\bone\bthree\right>\left[\btwo\bthree\right]^2$.  Among the ten remaining structures, those that are antisymmetric under $\btwo\leftrightarrow\bthree$ are arranged into the form 
\begin{equation}
\begin{aligned}
{\cal A}_B=&\left(b_1\left[\btwo\bthree\right]^2+b_2\left<\btwo\bthree\right>^2+b_3\left<\btwo\bthree\right>\left[\btwo\bthree\right] \right)\left(\left<\bone\btwo\right>\left[\bone\bthree\right]-\left<\bone\bthree\right>\left[\bone\btwo\right]\right)\,,
\end{aligned}
\label{allmassive}
\end{equation}
where $c_i$ are coupling coefficients.

A special case occurs for one massive $(m_\psi\neq0)$ and one massless $(m_\chi=0)$ Majorana spin-$3/2$ fermion coupling to a spin-1 boson, which, as we will see, corresponds to the spectrum of $N=2$ supersymmetry partially broken to $N=1$. If the spin-1 is massless, we denote $\psi_\mu$, $A_\mu$, $\chi_\mu$ by $\bone, 2, 3$. A basis for $SL(2,\mathbb{C})$ is simply $(\left|2\right>,\left|3\right> )$, and we obtain the amplitudes 
\begin{equation}
    \begin{aligned}
        &{\cal A}_A^{\left(  +1 ,  +\frac{3}{2} \right)}=a_1\left[\bone 2\right]\left[23\right]\left[\bone 3\right]^2 \,, 
        \\&{\cal A}_A^{ \left(-1 , -\frac{3}{2}  \right)}=a_2\left<\bone 2\right>\left<23\right>\left<\bone 3\right>^2  \,. 
    \end{aligned}
    \label{Apsichi}
\end{equation}

If the spin-$1$ boson is massive, we distinguish the case of unequal mass $m_\psi\neq m_B$ and equal mass $m_\psi=m_B$. The basis for the three-point amplitudes are different in these situations \cite{Arkani-Hamed:2017jhn}. For $m_\psi\neq m_B$, the $SL(2,\mathbb{C})$ space is spanned by the vectors  $(\left|3\right>,\frac{p_1}{m_\psi}\left|3\right])$, and after using the equations of motion we recover the following amplitudes
\begin{equation}
\begin{aligned}
  &  {\cal A}_B^{(+\frac{3}{2})}=\left[\bone 3\right]^2\left[\btwo 3\right]\left(\tilde{b}_1^+ \left[\bone \btwo\right]+\tilde{b}_2^+\left<\bone \btwo\right>\right)\,,\\  
  &{\cal A}_B^{(-\frac{3}{2})}=\left<\bone 3\right>^2\left<\btwo 3\right>\left(\tilde{b}_1^- \left<\bone \btwo\right>+\tilde{b}_2^-\left[\bone \btwo\right]\right)\,,
\end{aligned}\label{unequalBpsi}
\end{equation}
for $\chi_\mu $ with $\pm 3/2$ helicity and where $\bone $, $\btwo$, $3$ represent $\psi_\mu$, $B_\mu$, $\chi_\mu$ respectively.

On the other hand, for $m_\psi=m_B$, the basis in the above case is no longer independent and the three-point amplitude is instead given by the $x$-factor formula \eqref{nima-3}. Note that when the massless particle is a fermion, there are fractional powers in \eqref{nima-3}, but after using \eqref{xrelation}, the fractional powers will disappear. The first term in the $x$-factor expansion, corresponding to the minimal coupling, becomes
\begin{equation}\begin{aligned}
  &  {\cal A}_B^{(+\frac{3}{2})}= \tilde{b}_3^+\left(\left<\bone \varepsilon_3^+\btwo \right]+\left<\btwo \varepsilon_3^+\bone \right]\right)\left<\bone\btwo\right>\left[\btwo3\right]\,,\\  
  &{\cal A}_B^{(-\frac{3}{2})}= \tilde{b}_3^-\left(\left<\bone \varepsilon_3^-\btwo \right]+\left<\btwo \varepsilon_3^-\bone \right]\right)\left[\bone\btwo\right]\left<\btwo3\right>\,.
\end{aligned}\label{equalBpsi}
\end{equation}

We now comment on the dimension of the coupling coefficients. The three-point amplitude has mass dimension one, hence the couplings $g_i^{(\pm 2)}$, $g_i^{(+1)}$, $a_i$, $b_{i}$, $\tilde{b}^\pm_i$ must be normalized correspondingly by mass dimensions, which can be a generic mass scale or the mass of the particles. Typically, each polarization vector  contributes with   one inverse mass according to \eqref{eq:massivepolvect}. Furthermore, to account for interactions with gravity, we introduce the inverse Planck mass $\kappa\equiv 2/M_P$, where $M_P=2.4\times 10^{18}$~GeV, that can be used to normalize the couplings. 

\section{On-shell construction of $N=1$ supergravity}  \label{sec3}

Having set up the on-shell formalism, we now construct  spin-3/2 scattering amplitudes using recursion relations and identify possible spin-3/2 interactions. We will see that this uniquely leads to supergravity from the bottom up. We begin with just one Majorana spin-$3/2$ fermion, $\psi_\mu$, with mass $m_{3/2}\equiv m_\psi$ and systematically add new degrees of freedom. Our requirements are as follows:
\begin{itemize}
\item Interactions involving two spin-$3/2$ particles contain up to one derivative 
\item The three-point amplitudes have a smooth massless limit
\item The four-point amplitude satisfies the Ward identity in the massless limit
\end{itemize} 
The first requirement means that we do not aim to construct higher-derivative supergravity. The last two requirements smoothly connect the massive theory to a massless theory. More precisely,  the Ward identity involves the amplitude with all  \textit{transverse} external states, so that when one of the polarizations is replaced by momentum, the result must vanish in the massless limit. In addition, we impose \textit{unitarity} on the scattering amplitude so that there is an effective theory valid up to a cutoff $\Lambda\gg m_{3/2}$ that does not depend on the particle masses.

We first review the case with just one massive spin-3/2 particle.
Without additional degrees of freedom, the spin-3/2 fermion can self-interact via four-fermion, dimension six, contact terms. The spin-3/2 scattering amplitude is then determined by the coefficients of these interactions. Generically, the most divergent high-energy behavior arises from the scattering with all longitudinal states as external particles. In this case each polarization vector  grows as $\sim E^{3/2}$, and therefore the total four-point amplitude behaves as $E^6$, which could be   worse if there are derivative insertions.  Even if the interaction is 
scaled with $1/M^2$ with $M\gg m_{3/2}$, the cutoff of the amplitude will be of order $(m_{3/2} ^2M)^{1/3}$, so unitarity is violated at a low energy scale, if the spin-$3/2$ particle is light. 

The aim will be to raise the cutoff $\Lambda$ to be near the scale $M$ (which can be identified with the Planck scale, independent of particle masses) by systematically introducing interactions with other particles, so that new diagrams with boson exchanges will contribute to the final amplitude. 
In general, the effective field theory approach consists in adding all possible Lagrangian interactions and four-fermion contact terms, with generic coupling constants that may be constrained by symmetry arguments. However, it is tedious to cover all off-shell possibilities and the complexity increases with the particle spin. Alternatively, in the on-shell approach, the three-point amplitudes are dictated by little-group covariance and are independent of the Lagrangian \cite{Arkani-Hamed:2017jhn}. In this spirit, 
the effective three-point interactions involving massive spin-3/2 fermions were given in Ref.~\cite{Gherghetta:2024tob}.

Constructing massive amplitudes out of three-point building blocks is not necessarily automatic, as was first noticed for QED in \cite{Christensen:2022nja}.
This is because naively gluing three-point amplitudes leads to the so-called ``contact-term ambiguity". Since the three-point amplitude can be written in different forms that are equivalent on-shell, the product of them will not necessarily be the same.  One way to address this problem is to tune the coefficients of the contact terms to match the amplitude calculated by Feynman rules. This tuning by hand  then loses the bottom-up advantage of the on-shell formalism, since it still depends on the Lagrangian.

Ultimately, this problem is due to the lack of a good momentum shift, and the ``contact-term ambiguity'' is more properly speaking a ``boundary-term ambiguity'', since the boundary term in the recursion relation is unknown and probably non-vanishing. When the theory is constructible, the contact term should be fixed because the amplitude is unique. The all-line momentum shift introduced in Ref.~\cite{Ema:2024vww}, allows on-shell amplitudes to be constructed in QED \cite{Ema:2024vww}, the electroweak theory \cite{Ema:2024rss} and $N=1$ supergravity \cite{Gherghetta:2024tob}, without any ambiguity.  In these examples, the starting point was a well-defined theory containing specific particles and interactions
where constructibility ensures that the amplitudes match those that are computed from Feynman rules.

One can also invert the logic, and instead of constructing amplitudes from a well-defined theory, we can build up a set of possible interactions to construct the theory from the bottom up using the three-point amplitudes. 
By imposing the set of requirements listed at the beginning of this section, any theory satisfying them will be constructible under the momentum shift. The uniqueness of the amplitude then reveals the possible underlying effective field theories. As we will see, starting from a massive spin-$3/2$ particle and our requirements, eventually we are uniquely led to the interactions of (broken) supergravity.

\subsection{Adding spin 0 and spin 1}
\subsubsection{Scalar and pseudoscalar interactions}

Obviously, spin-$3/2$ alone cannot have a unitarity  cutoff near $M$. The simplest extension is then to couple the spin-$3/2$ fermion to a scalar field. From the three-point amplitudes \eqref{basis-phipsipsi}, there are two types of scalar interactions, either CP-even or CP-odd. The two on-shell interactions that satisfy our requirement of no higher derivatives are~\cite{Gherghetta:2024tob}
\begin{equation}
    \begin{aligned} 
    d_S  \psib^\mu\psi_\mu S+i\frac{d_P}{M} \varepsilon^{\mu\nu\alpha\beta } \partial_\mu P \psib_\nu \gamma_\alpha\psi_\beta\,,
     \end{aligned} 
     \label{PScouplingsd}
\end{equation}
where $d_{S,P}$ are dimensionless coefficients.
Since there is no gravity, so far, $M$ is a generic mass scale with $M\gg m_{3/2}$. By dimensional analysis, we conclude that the four-point spin-$3/2$ scattering amplitude is constructible under the ALT shift, for the couplings in \eqref{PScouplingsd}. There is no contact term generated by the construction, and the helicity $\left(\frac{1}{2},\frac{1}{2},\frac{1}{2},\frac{1}{2}\right)$ amplitude grows as $E^4$. In the center-of-mass frame with scattering angle $\theta$, the scattering amplitudes for $\left(\frac{1}{2},\frac{1}{2},\frac{1}{2},\pm \frac{1}{2}\right)$ helicity configurations, are given by\cite{Gherghetta:2024tob}
\begin{equation}
    \begin{aligned}
    &{\cal A}^{\left(\frac{1}{2},\frac{1}{2},\frac{1}{2}, \frac{1}{2}\right)}_{S}=-\frac{16}{9}d_S^2 \frac{E^4}{m_{3/2}^4}+\mathcal{O}(E^2/m^2)\,,
\\&{\cal A}^{\left(\frac{1}{2},\frac{1}{2},\frac{1}{2}, -\frac{1}{2}\right)}_S=\frac{8}{9}d_S^2 \sin2\theta \frac{E^3}{m_{3/2}^3}+\mathcal{O}(E/m)\,,
\end{aligned}
\label{scalarcom}
\end{equation}
for $S$ exchange and
\begin{equation}
    \begin{aligned}
    &{\cal A}^{\left(\frac{1}{2},\frac{1}{2},\frac{1}{2}, \frac{1}{2}\right)}_{P}= -\frac{64}{9}d_P^2 \frac{E^4}{m_{3/2}^2M^2}+\mathcal{O}(E^2/m^2)\,,
\\&{\cal A}^{\left(\frac{1}{2},\frac{1}{2},\frac{1}{2}, -\frac{1}{2}\right)}_P=0\,,
\end{aligned}
\label{pscalarcom}
\end{equation}
for $P$ exchange. With the scalar exchange only, the cutoff is of order $m_{3/2}$. For pseudoscalar exchange, the unitarity cutoff is of order $\sqrt{M m_{3/2}}\gg m_{3/2}$, which could be a valid effective theory for spin-$3/2$ without gravity. However, we are interested in an effective theory with a cutoff $\Lambda$ near $M$ (independent of particle masses), which cannot be achieved with scalar exchanges.

Finally, note that the Ward identity can be satisfied by assuming $d_S\propto \frac{m_{3/2}}{M}$, and this will be the case when we introduce gravity in section~\ref{sec:Majoranaspin2}.

\subsubsection{Massive spin-1 interaction}

As mentioned in Section~\ref{threepoints}, the three-point amplitude of a massless spin-$1$ boson interacting with a Majorana spin-$3/2$ fermion vanishes by Fermi statistics.   
Thus, the next possibility is to consider a massive spin-1 particle with mass, $m_B$. We will check whether unitarity and  the Ward identity of spin-3/2 scattering can be satisfied by adding this new particle.

The general three-point amplitudes are given in \eqref{allmassive}. We first  reduce this basis by imposing a smooth massless limit. The dimensionality of \eqref{allmassive} implies that  $b_{1,2,3}$ have  mass dimension $-3$, which should be some combination of particle masses in the absence of gravity.

It is straightforward to show that $b_1,b_2$ do not have a smooth massless limit. For the $b_1$ term, the $\left(1,\frac{1}{2},\frac{3}{2}\right)$ component gives
\begin{equation}
    \begin{aligned}
    b_1\left[\btwo\bthree\right]^2 \left(\left<\bone\btwo\right>\left[\bone\bthree\right]-\left<\bone\bthree\right>\left[\bone\btwo\right]\right)   \longrightarrow \frac{1}{\sqrt{3}}b_1 \left[23\right]^2\left[13\right]  \left<\eta_1 2\right>=\frac{1}{\sqrt{3}}b_1 m_B\left[23\right]\left[13\right]^2\,.
    \end{aligned}\label{c1amp}
\end{equation}
However, $b_1 m_B$ has a mass divergence $\propto 1/m^2$  
and hence this helicity configuration has no smooth massless limit. A similar conclusion is obtained for the $b_2$ term, by considering the $\left(0,-\frac{3}{2},-\frac{3}{2}\right)$ amplitude
\begin{equation}
    \begin{aligned}
         b_2\left<\btwo\bthree\right>^2 \left(\left<\bone\btwo\right>\left[\bone\bthree\right]-\left<\bone\bthree\right>\left[\bone\btwo\right]\right)\longrightarrow 
         \frac{1}{\sqrt{2}}b_2 m_{3/2}\left<23\right>^3\,.
    \end{aligned}\label{c2amp}
\end{equation}
Finally, for the $b_3$ term, the massless limits yield
\begin{equation}   b_3\left<\btwo\bthree\right>\left[\btwo\bthree\right]  \left(\left<\bone\btwo\right>\left[\bone\bthree\right]-\left<\bone\bthree\right>\left[\bone\btwo\right]\right) \rightarrow
    \left \{\begin{aligned}
    &\frac{1}{\sqrt{3}}b_3m_{3/2}m_B \frac{[12]^3 }{[13]}\quad \text{for} \quad \left(1,\frac{3}{2},-\frac{1}{2}\right)\\&- \frac{1}{\sqrt{3}}b_3m_{3/2}m_B\frac{[13]^3 }{[12]}
    \quad \text{for} \quad \left(1,-\frac{1}{2},\frac{3}{2}\right)\\&-\frac{2}{\sqrt{6}}b_3m_{3/2}^2\frac{[12][23]^2}{[13]}
    \quad \text{for} \quad \left(0,\frac{3}{2},\frac{1}{2}\right) \\& \frac{2}{\sqrt{6}}b_3m_{3/2}^2\frac{[13][23]^2}{[12]}
    \quad \text{for} \quad \left(0,\frac{1}{2},\frac{3}{2}\right)\\&-
    \frac{1}{\sqrt{3}}b_3m_{3/2}m_B \frac{\left<12\right>^3 }{\left<13\right>}\quad \text{for} \quad \left(-1,-\frac{3}{2},\frac{1}{2}\right)\\&  \frac{1}{\sqrt{3}}b_3m_{3/2}m_B\frac{\left<13\right>^3 }{\left<12\right>}
    \quad \text{for} \quad \left(-1, \frac{1}{2},-\frac{3}{2}\right)\\& \frac{2}{\sqrt{6}}b_3m_{3/2}^2\frac{\left<12\right>\left<23\right>^2}{\left<13\right>}
    \quad \text{for} \quad \left(0,-\frac{3}{2},-\frac{1}{2}\right) \\& -\frac{2}{\sqrt{6}}b_3m_{3/2}^2\frac{\left<13\right>\left<23\right>^2}{\left<12\right>}
    \quad \text{for} \quad \left(0,-\frac{1}{2},-\frac{3}{2}\right)\end{aligned}\right.
\label{masslessB}\end{equation} 
If $b_3\propto 1/m_{3/2}^3$ (or combinations with $m_B$), this amplitude is not smooth, either.\footnote{The $1/m^3$ singularity in the massless limit is related to the coupling of massive higher spins. More precisely, a three-point amplitude with massive spin $(s_1,s_2,s_3)$ is a combination of $2s_1$ $|\bone)$'s, $2s_2$ $|\btwo)$'s and $2s_3$ $|\bthree)$'s (the momenta are canceled by equations of motion). Therefore, the spinor structure has mass dimension $s_1+s_2+s_3$ which implies a coupling coefficient of mass dimension $1-(s_1+s_2+s_3)$.
Hence, larger spins $s_i$ leads to a more singular coupling in the massless limit.} Gravity can solve this problem by introducing a dimensionful coupling, so that $b_3\propto\kappa/m^2$, as we will see in Section~\ref{sec:Majoranaspin2}. Note, however, that the $b_{1},b_2$ amplitudes will remain divergent even with the graviton because the amplitudes \eqref{c1amp}-\eqref{c2amp} behave as $\kappa /m$ if $b_{1},b_2\propto \kappa /m^2$.    
Thus, anticipating introducing the graviton, we will use the $b_3$ term to construct the scattering amplitude, which is less divergent in the massless limit.
An on-shell, dimension-four operator corresponding to $b_3$ amplitude \eqref{masslessB} is 
\begin{equation}
    c_B \bar{\psi}^\mu \slashed{B}\gamma^5\psi_\mu\,.
  \label{threeptB}
\end{equation}
In the class of interactions equivalent on-shell to \eqref{threeptB} (assuming no derivatives in the interaction), the amplitude is always constructible by dimensional analysis\footnote{This argument is based on the fact that the interaction \eqref{threeptB} has no derivative. More precisely, the massive spin-1 propagator has a $p_\mu p_\nu/m^2$ term which may result in a 
$z^0$ scaling. However, one can explicitly check that the $p_\mu p_\nu/m^2$ piece in the amplitude separately satisfies the Ward identity in the massless limit, so the true $z$-scaling is up to $z^{-1}$.} (see Appendix~\ref{app:shift}). 

We now proceed to calculate the $\psi_1\psi_2\rightarrow\psi_3\psi_4$ scattering amplitude. 
Using Eq.~\eqref{four-point-fact}, with $s,t,u$ factorization channels, we obtain for the $\left(\frac{1}{2},\frac{1}{2},\frac{1}{2},\pm\frac{1}{2}\right)$ helicity configurations
\begin{equation}
\begin{aligned}
    &{\cal A}^{\left(\frac{1}{2},\frac{1}{2},\frac{1}{2}, \frac{1}{2}\right)}_{B}=-\frac{64}{9}c_B^2\frac{E^4}{m_{3/2}^2m_B^2}-\frac{32}{9}c_B^2\frac{E^4}{m_{3/2}^4}+\mathcal{O}(E^2/m^2)\,,
\\&{\cal A}^{\left(\frac{1}{2},\frac{1}{2},\frac{1}{2}, -\frac{1}{2}\right)}_B=\frac{8}{9}c_B^2\sin2\theta\frac{E^3}{m_{3/2}^3}+\mathcal{O}(E/m)\,.
\end{aligned}
\label{Bexpm}
\end{equation} 

So far, we have verified that a theory with a massive, Majorana spin-$3/2$ fermion and a complex scalar has an amplitude with an $E^4$ high-energy behavior,
hence violating unitarity near the  scale $\sqrt{M m_{3/2}}$, dependent on the particle mass. Furthermore, adding a massive spin-1 particle does not help to cancel the $E^4$ term. The massive spin-1 interaction has no smooth massless limit in the first place, unless $c_B\propto \frac{m}{M}$, and even if one constructs the amplitude with massive spin-1 exchange, the amplitude cannot be unitarized to order $E^2/M^2$, either.  

\subsection{Adding a spin-2 graviton}
\label{sec:Majoranaspin2}

Having exhausted all the the spin $\leq 1$ possibilities, the next simplest additions are adding a graviton and/or another Majorana spin-3/2 particle. The latter case will be discussed in Section~\ref{sec4}, so in this section we consider introducing a spin-2 graviton. An important consequence of gravity is a new dimensionful coupling  $\kappa=2/M_P$, that is used to normalize the interactions. We assume that all interactions contain $\kappa$, and ignore larger powers of $\kappa$ that are Planck suppressed, higher-order effects. As a result, the previous issues related to massless limits are solved. For the scalars, we rewrite the couplings \eqref{PScouplingsd} as
\begin{equation}
    \begin{aligned} c_S \kappa m_{3/2}  \psib^\mu\psi_\mu S+i c_P\kappa  \,\varepsilon^{\mu\nu\alpha\beta } \partial_\mu P \psib_\nu \gamma_\alpha\psi_\beta \,,
     \end{aligned} 
     \label{PScouplings}
\end{equation}
with $d_S=c_S \kappa m_{3/2}$, $d_P=c_P \kappa M $. Now $d_P$ is fixed, while  $d_S$ is  no longer fixed and vanishes in the $m_{3/2}\rightarrow0$ limit. The transverse amplitudes, with one polarization replaced by momentum, are all proportional to $d_S^2$, $d_P^2 m_{3/2}^2$ in the massless limit. When $d_S^2$, $d_P^2m_{3/2}^2$ vanish in this limit, the Ward identity is then realized. 

As for the $B_\mu$ coupling, the least divergent piece, $b_3$, now has a smooth massless limit when $b_3\propto \frac{\kappa}{m_{3/2}^2}$, whereas   $b_1$ and $b_2$ are still singular in this limit.  Notice that the  massless amplitudes in \eqref{masslessB} are composed of helicities $0$, $\pm 1/2$ and $\pm 3/2$, which point  to the coexistence of a Higgs and a super-Higgs mechanisms for spin-1 and $3/2$. In the high-energy limit, $E\gg m$, they can respectively, be understood as the coupling of a Nambu-Goldstone boson, a Goldstino and the massless gravitino.  One can start with the massless three-point amplitudes:
\begin{center}
\begin{tabular}{ccccc} 
    \adjustbox{valign=m}{\includegraphics[width=0.2\textwidth]{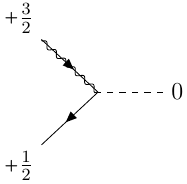}}  & $=\kappa\frac{[13][23]^2}{[12]}$&\qquad & 
    \adjustbox{valign=m}{\includegraphics[width=0.2\textwidth]{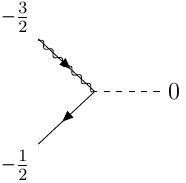}} &$ 
    =\kappa\frac{\left<12\right>\left<23\right>^2}{\left<13\right>}$ 
\end{tabular}
\end{center}
They have an identical dimensionful coupling which is taken to be $\kappa$. In the IR limit, they combine into a single object~\cite{Arkani-Hamed:2017jhn} which is the $b_3$ amplitude \eqref{masslessB}. Conversely, given the massive amplitude and taking the massless limit, the amplitudes that survive are the two given above. Again, these limits are justified thanks to the introduction of gravity with a dimensionful coupling $\kappa$. 

The other issue was unitarity. Given the new mass scale $M_P$, we can aim to push the perturbative unitarity bound to be near $M_P$, which is much larger than any particle mass scales. Since the four-point amplitude has a factor $\kappa^2$, the high-energy behavior can be up to $E^2$. To investigate the unitarity of the full amplitude, we shall compute the three-point amplitudes with the graviton coupling.

In \cite{Gherghetta:2024tob}, we have shown that the three-point amplitude with a smooth massless limit and no higher derivative turns out to be the minimal coupling (corresponding to the ${ g}_0^{(\pm2)}$ terms in $\mathcal{A}_h^{(\pm2)}$). This amplitude is equivalent on-shell  to the interaction $c_h\kappa h_{\mu\nu}T_{3/2}^{\mu\nu}$, with  $T_{3/2}^{\mu\nu}$ being the energy-momentum tensor
\begin{equation} 
c_h\kappa h_{\mu\nu} \left. T_{3/2}^{\mu\nu}\right|_\text{on-shell}= i\frac{c_h}{2}\kappa h_{\mu\nu}\left(\psib^\rho \gamma^\mu \partial_\rho \psi^\nu -\partial_\rho \psib^\mu \gamma^\nu \psi^\rho -\frac{1}{2}\psib_\rho \gamma^\mu  {\partial^\nu}\psi^\rho+\frac{1}{2}{\partial^\nu}\psib_\rho \gamma^\mu    \psi^\rho\right)\,,
\label{mincouponshell}
\end{equation}
where $c_h$ is a dimensionless coefficient. By dimensional analysis and the Ward identity (see Appendix~\ref{app:shift}), the four-point amplitude is constructible under the ALT shift. There are three factorization channels ($s,t,u$), and the three-point amplitude in the vector notation is
\begin{equation}
\begin{aligned}
    &V^{h\psi_1\psi_2}(p_1,p_2,p_I) \, 
    \\&=c_h\kappa\,\varepsilon^{\mu\nu}(p_I)
    \left[\frac{1}{2}\bar{v}_{2\alpha} \gamma_{(\mu}(p_1-p_2)_{\nu)}u_{1}^\alpha-\Bar{v}_{2\alpha }\gamma_{(\mu}(p_1-p_2)^\alpha u_{1\nu)}-\Bar{v}_{2(\nu}\gamma_{\mu)}(p_1-p_2)_\alpha u_{1}^\alpha\right]\,,
\end{aligned}
\label{Vhpsipsiexp}
\end{equation} 
where $\varepsilon^{\mu\nu}$ is the graviton polarization vector, with symmetrized $\mu\nu$ indices, and $\bar{v}_\mu\equiv \bar{v}\varepsilon_\mu$, $u_\mu\equiv u\varepsilon_\mu$ are the spin-$3/2$ polarization vectors with $u,\bar{v}$ the Dirac spinors. Since \eqref{Vhpsipsiexp} has only one momentum insertion, the momentum shift deforms the three-point amplitude by a term linear in $z$. The product of two deformed amplitudes is then up to $z^2$ order. Using the residue formula, the quadratic $z^2$ term gives rise to contact terms\cite{Gherghetta:2024tob}
\begin{equation}
\begin{aligned}
 {\cal A}_\text{contact}= \frac{1}{8}c_h^2\kappa^2
    &\left[\bar{v}_1 ^\mu \gamma^\nu  u _2^\alpha \bar{v}_{3\mu }\gamma_\nu  u_{4\alpha} -\bar{v}_1 ^\mu \gamma^\nu  u _2^\alpha \bar{v}_{3\alpha }\gamma_\nu  u_{4\mu}+\bar{v}_1 ^\mu \gamma^\alpha  u _2^\nu \bar{v}_{3\alpha }\gamma_\mu  u_{4\nu}\right.\\
    &\left.+\,\bar{v}_1 ^\mu \gamma^\alpha  u _2^\nu \bar{v}_{3\mu }\gamma_\nu  u_{4\alpha} -\bar{v}_1 ^\mu \gamma^\nu  u _2^\alpha \bar{v}_{3\nu }\gamma_\alpha  u_{4\mu}-\bar{v}_1 ^\alpha \gamma^\nu  u _2^\mu \bar{v}_{3\mu }\gamma_\alpha  u_{4\nu}\right] 
    -(1\leftrightarrow3)-(1\leftrightarrow4)\,.
\end{aligned}    
\label{psicontactterm}
\end{equation}
These contact terms are \textit{uniquely fixed} by the on-shell constructive method, and are related to the three-point interactions via the Ward identity. When $c_h=\frac{1}{2}$, corresponding to the gravitational minimal coupling ($\frac{\kappa}{2}  h_{\mu\nu} T^{\mu\nu}_{3/2}$), these contact terms are precisely those in  $N=1$ minimal supergravity.  
The same effect arises with photon exchange, and we will consider this case in more detail in Section~\ref{sec4}.

In the center-of-mass frame, the graviton exchange yields
\begin{equation}
    \begin{aligned}
    &{\cal A}^{\left(\frac{1}{2},\frac{1}{2},\frac{1}{2}, \frac{1}{2}\right)}_{h}= \frac{16}{3}c_h^2\frac{\kappa^2E^4}{m_{3/2}^2}+ \mathcal{O}(\kappa^2E^2)\,,
\\&{\cal A}^{\left(\frac{1}{2},\frac{1}{2},\frac{1}{2}, -\frac{1}{2}\right)}_h=-\frac{4}{3}c_h^2\sin2\theta\,\frac{\kappa^2E^3}{m_{3/2}}+\mathcal{O}(\kappa^2m_{3/2}E)\,,
\end{aligned}
\label{gravexpm}
\end{equation}
where the $E^6$ and $E^5$ terms are canceled between the contact and propagator pole terms.  We now fix $c_h=\frac{1}{2}$. Adding up \eqref{scalarcom}, \eqref{pscalarcom}, \eqref{Bexpm} and \eqref{gravexpm}, requiring unitarity (up to $E^2$ terms) then results in 
\begin{equation}
    c_B^2=- \left(c_S^2-\frac{3}{8}\right) \kappa^2 m_{3/2}^2,\qquad c_P^2=\frac{1}{4}c_S^2 -\frac{c_B^2}{\kappa^2 m_B^2}\,.
    \label{conditionmA}
\end{equation}
Imposing the above relations, the total amplitudes become
\begin{equation}
\begin{aligned}
 {\cal A}_{\text{total}}^{\left(\frac{1}{2},\frac{1}{2},\frac{1}{2}, \frac{1}{2}\right)}=&\frac{\kappa^2 E^2 }{144}\csc\theta^2 \left[ 384  + 350 c_S^2  - 16 ( 8 c_S^2-3 ) \mu_B^2 
 - 32 c_S^2 \mu_S^2\right.  \\&\left. + \,(192 - 286 c_S^2  + 16 ( 8 c_S^2-3 ) \mu_B^2 + 32 c_S^2 \mu_S^2) \cos 2\theta \right.  \\&\left.
 -\,104 c_S^2  \cos 4 \theta + 29 c_S^2  \cos 6 \theta  + 10 c_S^2 \cos 8 \theta  + c_S^2 \cos 10 \theta 
 \right]+\mathcal{O}(\kappa^2 m_{3/2}^2)\,,\\
 {\cal A}_{\text{total}}^{\left(\frac{1}{2},\frac{1}{2},\frac{1}{2}, -\frac{1}{2}\right)}=&-\frac{1}{8}\kappa^2 m_{3/2}E\csc \theta\left(15 \cos \theta +\cos 3\theta\right)+\mathcal{O}\left( {\kappa^2m_{3/2}^3}/{E}\right)\,,
\end{aligned}
\label{eq:gravpmhalf}
\end{equation}
where $\mu_{B,S}^2\equiv m_{B,S}^2/m_{3/2}^2$. Clearly, the unitarity cutoff, determined by the most divergent helicity amplitude in \eqref{eq:gravpmhalf}, is now near the Planck scale. Note that Eq.~\eqref{conditionmA} has an apparent singularity in the $c_P$ coupling when $m_B\rightarrow 0$. In fact, since $c_B$ is dimensionless and $b_3\propto \frac{c_B}{m_{3/2}^2 m_B}\propto\frac{\kappa}{m_{3/2}^2}$ for the amplitude to have a smooth massless limit, we can derive the condition $c_B \propto \kappa m_B$. According to the first equation in \eqref{conditionmA}, we then have $m_B\propto m_{3/2}$ so that none of the couplings $c_S$, $c_P$, $c_B$ has a mass singularity. 

The search for an effective theory of spin-3/2 scattering valid up to $\Lambda\gg m_{3/2}$ 
is closely related to supersymmetry breaking mechanisms, and this can be seen from our bottom-up approach. Indeed, when there is no massive spin-1 coupling, i.e.~$c_B=0$, we obtain $c_S^2=3/8$ and $c_P^2=c_S^2/4$. These on-shell coupling values precisely match those in the Polonyi model \cite{Gherghetta:2024tob}, where supersymmetry is spontaneously broken by coupling the spin-3/2 fermion to a chiral multiplet containing a spin-1/2 Weyl fermion (Goldstino) and a complex scalar field. Notice also, that both the scalar and pseudoscalar particles of the complex scalar field are needed for unitarity at $E^2$ order.

More generally, when $c_B\neq 0$, the theory can be matched to $N=1$ supergravity spontaneously broken to $N=0$ by both $F$- and $D$-terms.\footnote{The couplings $c_B$, $c_P$, $c_S$ are functions of the Kähler potential, superpotential, moment map, and gauge kinetic function \cite{Antoniadis:2023cdw}.}   The $U(1)$ gauge boson in the vector multiplet acquires a mass via the Higgs mechanism, absorbing one scalar degree of freedom from the chiral multiplet\footnote{In the unitary gauge, the remaining scalar degree of freedom has both CP-even and CP-odd couplings to the gravitino, of the type Eq.~\eqref{PScouplings}.} containing the Goldstino.  The axial vector coupling \eqref{threeptB} is also equivalent on-shell to the coupling in $N=1$ supergravity \cite{Freedman:1976uk}\footnote{Similar axial-vector couplings exist in global supersymmetry with $D$-term breaking \cite{Fayet:1977yc,Fayet:1979sa}, while for amplitudes involving gravitinos, the role played by the axial coupling has been studied in \cite{Fayet:1979yb}.}, which arises from the Kähler connection in the gravitino covariant derivative. 
Unitarity to order $E^2$ has been explicitly checked using the supergravity Lagrangian and Feynman rules in Ref.~\cite{Antoniadis:2023cdw}. Thus, in our bottom-up approach we are uniquely led to the superHiggs and Higgs mechanism in order to obtain an interacting effective theory that is valid up to the Planck scale, and matches known supersymmetry breaking results in $N=1$ supergravity theories.

Having derived a set of three-point amplitudes satisfying our initial requirements and leading to a unitary scattering amplitude to ${\cal O}(\kappa^2 E^2)$, it is now interesting to recover the interactions in the high-energy limit. 
The three-point interactions leading to a spontaneously broken $N=1$ supergravity, and their UV (unbroken) limits are summarized as follows: 
\begin{center}
    \begin{tabular}{c|c}
IR (massive)   & UV  (massless)\\ \hline
 $\psi_\mu\psi_\nu h_{\alpha\beta}$    & $ \psi_{T\mu}\psi_{T\nu}  h_{\alpha\beta} $, $ \psi_{L}\psi_{L}  h_{\alpha\beta} $ \\  
 $\psi_\mu\psi_\nu\phi$    &   $ \psi_{T\mu }\psi_{L}  \phi_H $  \\  
 $\psi_\mu\psi_\nu B_\mu$    & $ \psi_{T\mu }\psi_{L}  A_\alpha $, $ \psi_{T\mu }\psi_{L}  \phi_E $  \\
  $B_\mu B_\nu \phi$  &  $A_\mu \phi_H\phi_E$
\end{tabular}
\end{center}
where $\phi$ represents a scalar that can have CP-even ($S$) and/or CP-odd  ($P$) interactions with the massive spin-$3/2$ fermion. The massless states include 
$\phi_H$, a Higgs-like state that remains in the spectrum, and $\phi_E$ that is eaten by the photon in the IR,
while $\psi_{T\mu}$, $\psi_L$ are the transverse (helicity $\pm 3/2$) and longitudinal (helicity $\pm 1/2$) components of the massive spin-$3/2$ particle in the high-energy limit, corresponding to a massless gravitino and Goldstino, respectively, in a supersymmetric theory. 
The second column of the first row involves the gravitational interactions for the massless particles. In the second row, the gravitino-Goldstino-scalar UV interaction is packaged into the IR massive $\psi_\mu\psi_\nu\phi$ coupling, which reflects the superHiggs mechanism \cite{Arkani-Hamed:2017jhn}: if we regard $\phi_H$ as the fluctuation of a field $H=\left<H\right>+\phi_H$, then the vacuum expectation value (VEV) part gives a mass term for the gravitino.  A similar scalar interaction exists in the UV in the third row, except that the scalar $\phi_E$ is eaten by
the massless photon to become a massive photon in the IR. This is the case with $F$- and $D$-term supersymmetry breaking. The Higgs mechanism associated with $B_\mu$ can be seen in the last row. One can perform a similar calculation with massive spin-$1$ scattering and show that the scalar exchange unitarizes the amplitude \cite{Ema:2024rss}.

\section{$N=2$ supergravity from a charged spin-3/2 fermion}
\label{sec4}

Previously, we have considered a single Majorana spin-3/2 fermion $\psi_\mu$, and shown that gravity is a necessary ingredient to obtain an effective theory valid to ${\cal O}(\kappa^2 E^2)$. Our requirements have led to the theory of broken $N=1$ supergravity. We now extend the particle content with another Majorana spin-$3/2$ state and consider the Dirac spinor $(\chi_\mu,\psi_\mu)$ charged under a $U(1)$ symmetry with a massless gauge boson. We will show that the Ward identity of the four-point amplitude $\chi_\mu\psi_\mu\rightarrow\chi_\mu\psi_\mu$ requires the introduction of a graviton whose coupling is consistent with $N=2$ minimal supergravity.\footnote{We can also start with $(\chi_\mu,\psi_\mu)$ coupling to a graviton,  then the Ward identity requires an additional coupling to spin 1, that can be either massless or massive. The latter case is discussed in Section~\ref{sec5}.  Alternatively, we can first couple identical spin-$3/2$ fermions to scalars and a massive spin-$1$ boson, then the same reasoning in Section~\ref{sec3} implies the existence of a graviton, which then leads to the coupling of $(\chi_\mu,\psi_\mu)$ to a massless photon.}

We consider the same assumptions as in Section~\ref{sec3}: (i) the three-point amplitude has a smooth massless limit, (ii) the four-point amplitude satisfies the Ward identity in the massless limit, (iii) the interactions contain up to one derivative. The last two assumptions ensure that the four-point amplitude is constructible under the ALT shift, which allows us to use the on-shell three-point amplitudes as building blocks. Furthermore, we impose unitarity.

\subsection{Spin-3/2 coupled to a photon}
\label{sec41} 

As we have argued in Section~\ref{threepoints}, a single Majorana spin-$3/2$ fermion cannot couple to a $U(1)$ gauge boson, so the minimal extension to make such a coupling possible is to introduce another Majorana spin-$3/2$ fermion, which together form a Dirac fermion. The three-point amplitudes dictated by little-group covariance are given by \eqref{gravphoton}. Since we are restricting to dimension $\leq5$ operators (up to one derivative insertion), we will not consider the $g_3^{(+1)}$ term which corresponds to higher-dimensional couplings. The remaining $g_{0,1,2}^{(+1)}$ terms will be  translated  into the vector notation, so that we can easily connect with Lagrangian interactions later. 

We denote the two Majorana fermions by $\chi_\mu$, $\psi_\mu$, and assume that they have equal mass $m_\chi=m_\psi\equiv m_{3/2} $ and form a Dirac fermion. The massless gauge boson, $A_\mu$ has polarization vector $\varepsilon_\mu$.  Restricting to dimension $\leq 5$ couplings, we can write down the following on-shell couplings from the $g_{0,1,2}^{(+1)}$ terms in \eqref{gravphoton}
\begin{equation}
   i e\bar{\chi}_\mu \slashed{A}\psi^\mu +\frac{l_1}{M}\bar{\chi}_\mu F^{\mu\nu}\psi_\nu +\frac{l_2}{M}\bar{\chi}_{\mu } F^{\al\beta}\gamma_{\al\beta}\psi^\mu\,,
\label{vector3ptphoton}
\end{equation}
where $M$ is a UV mass scale and $e$ is the $U(1)$ coupling which arises from the gauge covariant derivative $\mathcal{D}_\mu=\partial_\mu +ieA_\mu $.  To avoid further mass divergences, we will consider $M$ to be \textit{fixed} in the massless limit $m_{3/2}\rightarrow0$. The on-shell couplings of the form \eqref{vector3ptphoton} were also recently used to compute Compton amplitudes \cite{Ema:2025qgd}. The on-shell 3-point amplitudes are
\begin{equation}
    {\cal A}_A=J_\mu\varepsilon^\mu\,,
    \label{a3vec}
\end{equation}
with\begin{equation}
    J_\mu=e\,\bar{v}_{2\nu }\gamma_\mu u_1^\nu+\frac{l_1}{M}\left( p_1^\nu \bar{v}_{2\nu}u_{1\mu }-p_2^\nu \bar{v}_{2\mu}u_{1\nu}\right)-2\frac{l_2}{M}
    \left( p_1^\nu + p_2^\nu\right) \bar{v}_{2\al}\gamma_{\mu\nu}u_{1 }^\al\,, 
    \label{currentJ}
\end{equation}
where $\bar{v}_2, u_1$ are spin-3/2 polarization vectors for $\bar{\chi}_2$, $\psi_1$. In the spinor notation, where the two Majorana fermions are represented by $\bone$, $\btwo$ and $3$ corresponds to the (massless) photon, \eqref{a3vec} then becomes
\begin{equation}\begin{aligned}{\cal A}_A^{(+1)}
=\frac{1}{2m_{3/2}M}\left(\left<\btwo\varepsilon^+_3\bone\right]+\left<\bone\varepsilon^+_3\btwo\right]\right)
\left[l_1  \left<\bone\btwo\right>^2 -(l_1+4l_2)\left[\bone\btwo\right]^2 +\left(  4l_2  -\frac{2eM}{m_{3/2}} \right)\left<\bone\btwo\right>\left[\bone\btwo\right]  \right]\,.
\end{aligned} 
\label{threeptcompo}
\end{equation} 
For ${\cal A}_A^{(-1)}$ with  negative helicity, one replaces $\varepsilon^+_3$ by $\varepsilon_3^-$, and angles and squares are exchanged. The minimal coupling, defined as the first term in the $x$-factor expansion \eqref{nima-3}, corresponds to \begin{equation}
    l_1+4l_2=0,\quad l_2=\frac{e M}{2m_{3/2}}\,.
    \label{minimalcond}
\end{equation}
It has been shown that the minimal coupling amplitude with $l_1 = -2 e M/m_{3/2}$, is equivalent to a gyromagnetic ratio $g=2$ \cite{Chung:2018kqs}, which is an important property for elementary particles of arbitrary spin \cite{Ferrara:1992yc}.

In addition, one can check that the requirement of a smooth massless limit coincides with this minimal choice. For example, in the  massless limit ($m_{3/2}\rightarrow0$), the $\left(\frac{3}{2},-\frac{1}{2},1\right)$ helicity configuration of  ${\cal A}_A$ becomes
\begin{equation}
    {\cal A}_A^{\left(\frac{3}{2},-\frac{1}{2},1\right)}\propto \left(4l_2 -\frac{2e M}{m_{3/2}}\right)\frac{1}{m_{3/2}}\frac{[13]^3}{[23]}\,,
\end{equation}
while the $\left(\frac{3}{2},\frac{1}{2},1\right)$ helicity amplitude behaves as
\begin{equation}
    {\cal A}_A^{\left(\frac{3}{2},\frac{1}{2},1\right)}\propto \frac{1}{m_{3/2}M}\left(l_1+4l_2 \right)[13]^2[12],
\end{equation}
which implies that the last two terms in \eqref{threeptcompo} diverge in the  massless limit.
Hence, setting them to zero yields the minimal coupling relations \eqref{minimalcond}. 

\subsubsection{On-shell construction of the four-point amplitude}
For the on-shell construction of the four-point amplitude from the building blocks \eqref{vector3ptphoton}, we do not need to assume \eqref{minimalcond}, as constructibility a priori only requires the Ward identity and no higher derivatives.
We assume for now, generic couplings $l_1,l_2,e$ which will be later constrained by the Ward identity.  
 
Consider the four-point amplitude $\psi_1\chi_2\rightarrow \chi_3\psi_4$.  
As explained in Appendix~\ref{app:shift}, we can first construct amplitudes with  all external particles \textit{transverse}, and then amplitudes with longitudinal states are obtained by applying spin-raising and lowering operators. With photon exchange, there are two factorization channels,  $s$ and $t$:
\begin{figure}[H]
  \centering 
  \begin{minipage}[t]{0.4\textwidth}
    \centering
    \raisebox{-42pt}{\includegraphics[height=2.8cm]{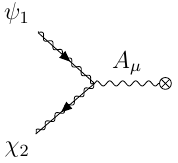}}
\raisebox{-43pt}{\includegraphics[height=2.8cm]{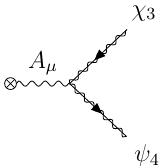}}
  \end{minipage}
  \hspace{0.15\textwidth} 
  \begin{minipage}[t]{0.2\textwidth}
    \centering
\hspace{-3pt}\includegraphics[width=0.8\linewidth]{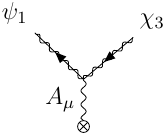} \\[1pt]
    \includegraphics[width=0.8\linewidth]{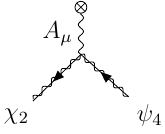}
  \end{minipage}
\end{figure}

Constructibility under the ALT shift follows from dimensional analysis.  The large-$z$ behavior is $z^0=z^2/z^2$, where $z^2$ is from the derivative at each vertex and $1/z^2$ is from the propagator. Furthermore, we assume that the four-point amplitude satisfies the Ward identity, i.e., it vanishes in the massless limit when one of the external polarizations is replaced by the momentum. This improves the large-$z$ behavior to $1/z$ and hence makes the amplitude constructible. From \eqref{four-point-fact}, the four-point amplitude can be rewritten as
\begin{equation}
    \begin{aligned}
        {\cal A}_4&=-\sum _{s,t}\sum _{z=z_I}\text{Res} \left[\hat{J}^\mu\hat{J}_\mu \frac{1}{z}\frac{1}{\hat{p}_I^2 -m_I^2}\right]\,,\\&=\sum _{s,t}\frac{1}{p_I^2}\frac{1}{z^+-z^-}\left[z^+\hat{J}^\mu(z^-)\hat{J}_\mu(z^-)-z^-\hat{J}^\mu(z^+)\hat{J}_\mu(z^+) \right]\,,
    \end{aligned}
    \label{four-point-construction}
\end{equation}
where  $p_I =p_1+p_2$ for the $s$-channel and $p_I =p_1+p_3$ for the $t$-channel. After the ALT shift $p_i\rightarrow \hat{p}_i=p_i+z r_i$, $\hat{J}^\mu$ depends linearly on $z$, so we can write\begin{equation}
    \hat{J}^\mu(z)\hat{J}_\mu(z )
=f_0+zf_1+z^2 f_2\,.
\end{equation}
Using \eqref{four-point-fact} and the useful relation $z^+z^-=\frac{p_I^2}{2r_i\cdot r_j}$, we  notice that $f_0$ gives the propagator pole term, $f_1$ has no contribution to the residue, and $f_2$ generates a contact term. This is similar to the gravitino contact term from graviton exchange considered in Section~\ref{sec3}.
The four-point amplitude is then given by
\begin{equation}
    {\cal A}_4=\frac{1}{p_I^2}f_0-\frac{1}{2r_1\cdot r_2}f_2-(2\leftrightarrow 3)\,.
    \label{A4general}
\end{equation}
For the $s$-channel we obtain the expressions
\begin{equation}
    \begin{aligned}
       f_0=&\left[e\, \bar{v}_{2\nu }\gamma_\mu u_1^\nu+\frac{l_1}{M}\left( p_1^\nu \bar{v}_{2\nu}u_{1\mu }-p_2^\nu \bar{v}_{2\mu}u_{1\nu}\right)
       -2\frac{l_2}{M}
        \left( p_1^\nu+ p_2^\nu\right)\bar{v}_{2\al}\gamma_{\mu\nu}u_{1 }^\al 
       \right] 
       \times \left(2\rightarrow3,1\rightarrow4\right)\,,\\ 
       f_2=&\left[ \frac{l_1}{M}\left( r_1^\nu \bar{v}_{2\nu}u_{1\mu }-r_2^\nu \bar{v}_{2\mu}u_{1\nu}\right)
       -2\frac{l_2}{M}
\left(r_1^\nu+r_2^\nu\right)\bar{v}_{2\al}\gamma_{\mu\nu}u_{1 }^\al
       \right] 
       \times \left(2\rightarrow3,1\rightarrow4\right)\,.
    \end{aligned}
    \label{eq:f0f2}
\end{equation} 
The second term in \eqref{A4general} depends on the shifted momenta and therefore is not little-group covariant. This term can be simplified by noting that under the ALT shift, the shifted momenta $r_i$ are proportional to the transverse polarization vectors. Writing $\varepsilon_i=c_i r_i$, where $c_i$ are proportionality constants, the second term in \eqref{A4general} becomes 
\begin{equation}
-\frac{f_2}{2 r_1\cdot r_2}=c_1c_2c_3c_4\frac{1}{M^2} (\bar{v}_2 u_1)(\bar{v}_3 u_4)(l_1+2l_2)^2 (r_1\cdot r_2)(r_1-r_2)\cdot r_3\,.
\label{firstfrr}
\end{equation}
Because of the little-group covariance of the amplitude, the result must be independent of the shifted momenta. The expression \eqref{firstfrr} can be rewritten in terms of  $r_1,r_2,r_3,r_4$ using the relation \eqref{rrelation1} in Appendix~\ref{app:contact}, so that  all the $c_i$ can be combined with the $r_i$ to form $\varepsilon_i$
and hence Eq.~\eqref{firstfrr} becomes manifestly independent of the shift.

As shown in Appendix~\ref{app:contact}, there are \textit{two} possible contact terms that give the same on-shell amplitude in \eqref{firstfrr}
\begin{equation}  
\xi_1 \left[(\bar{v}_2^ \mu  u_1^\nu)(\bar{v}_{3\mu} u_{4\nu})-(\bar{v}_2^\mu u_{1 \nu})(\bar{v}_3^\nu u_{4\mu})\right]-\left( 2\leftrightarrow 3\right)\,,
\label{contactg1}
\end{equation}
and
\begin{equation} 
\xi_2\left[ (\bar{v}_2^ \mu  u_1^\nu) (\bar{v}_{3}^\alpha \gamma_{\mu\nu} u_{4\al})+(\bar{v}_2^ \al  \gamma_{\mu\nu}u_{1\al}) (\bar{v}_{3}^\mu  u_{4}^\nu)\right] -\left( 2\leftrightarrow 3\right)\,,
\label{contactg2}
\end{equation}
with $\xi_1+2\xi_2=\frac{(l_1+2l_2)^2}{M^2}$. They are symmetric under the exchange $(2,1 \leftrightarrow3,4)$ and antisymmetric under $\left( 2\leftrightarrow 3\right)$. It may seem that the momentum shift does not uniquely fix the contact terms, because we have the freedom to choose $\xi_1$, $\xi_2$. However, constructibility under the ALT shift relies on the Ward identity, that must be satisfied by the resulting amplitude, so there are additional constraints on the contact terms.

To determine these constraints, recall that for an amplitude of the form ${\cal A}_4=\bar{v}_2^\mu \mathcal{M}_\mu\equiv \varepsilon_2^\mu\bar{v}_2 \mathcal{M}_\mu$, the Ward identity in the massless limit ($m_{\psi},m_{\chi}\rightarrow0$) imposes the constraint
\begin{equation}
  \tilde{\cal A}_4\equiv  p_2^\mu \bar{v}_2\mathcal{M}_\mu = 0\,,
  \label{eq:A4Wardid}
\end{equation}
where the polarization vector has been replaced by $p_2^\mu$.  
The first term in \eqref{currentJ} has no derivative and  its contribution to  
$\tilde{\cal A}_4$ cannot be canceled by the derivative interactions in the massless limit without additional constraints. Notice that the equations of motion imply
 \begin{equation}
    -2\frac{l_2}{M}
    \left( p_1^\nu + p_2^\nu \right)\bar{v}_{2\al}\gamma_{\mu\nu}u_{1 }^\al
    =-4\frac{l_2}{M}m_{3/2}\bar{v}_{2\al}\gamma_\mu u_1^\alpha -2\frac{l_2}{M}(p_{2\mu }-p_{1\mu})\bar{v}_2^\al u_{1\al}\,.
    \label{eq:eomrel}
\end{equation}
Namely,  the third term in \eqref{currentJ} can be equivalently written as in \eqref{eq:eomrel}, which contains a non-derivative interaction like the first term of \eqref{currentJ}.  Thus, for the non-derivative term to cancel in \eqref{eq:A4Wardid}, one requires the coupling, $e\propto \frac{l_2 }{M}m_{3/2}$, so that the massless limit is equivalent to the $e\rightarrow0$
limit, with $M$ fixed. The proportionality coefficient between the mass and coupling $e$ can be fixed by requiring a smooth massless limit, giving the same relations as in Eq.~\eqref{minimalcond}.
 
Having canceled the contribution from the first term in \eqref{currentJ}, we next check the Ward identity for the $l_1$ and $l_2$ terms. 
In the massless limit  ($p_i\rightarrow k_i$, $k_i^2=0$),\footnote{Note that $k_i$ is different from the shifted momentum $r_i$, since $k_i$ is not necessarily proportional to the polarization vector $\varepsilon_i$.} the propagator pole part of $\tilde{\cal A}_4$ in \eqref{A4general} arises from $f_0/p_I^2$. Using \eqref{eq:f0f2} we obtain
\begin{equation}
    \begin{aligned}
    & \frac{l_1^2}{2M^2}(\bar{v}_2 u_{1\mu})\left[(k_4\cdot \bar{v}_3)u_4^\mu -\bar{v}_3^\mu (k_3\cdot u_4)\right] -\frac{l_1l_2}{M^2}\left[\bar{v}_2 (u_1\cdot k_3)-\bar{v}_2 (u_1\cdot k_4)\right](\bar{v}_3\cdot u_4)  \\ &+ \frac{1}{2M^2}(l_1+4l_2)\frac{1}{k_1\cdot k_2}\bar{v}_2(k_2\cdot u_1)\left[l_1 (k_2\cdot \bar{v}_3)(k_3\cdot u_4)-l_1 (k_4 \cdot \bar{v}_3)(k_2\cdot u_4)\right.\\&\hspace{5cm}\left.+2l_2 (k_1 \cdot k_4)(\bar{v}_3\cdot u_4)-2l_2 (k_1 \cdot k_3)(\bar{v}_3\cdot u_4)\right]
    -\left( 1\leftrightarrow 4\right)\,.
    \end{aligned}
    \label{propwi}
\end{equation}
The above expression must be canceled by the contact term part in $\tilde{\cal A}_4$, but the latter does not have a denominator $1/(k_1\cdot k_2)$, so we have to impose $l_1+4l_2=0$ for the second and third lines in \eqref{propwi} to vanish. Under this condition, the on-shell couplings can be equivalently rewritten as $\psib_\mu (F^{\mu\nu} -i\gamma^5 \widetilde{F}^{\mu\nu})\psi_\nu $, which are the same as the dipole couplings appearing  in $N=2$ supergravity \cite{Ferrara:1976fu,Freedman:1976aw,Ferrara:1976iq}, with the minimal Lagrangian given in Appendix~\ref{app:sugra}. 

Now it remains to cancel only the first line in \eqref{propwi}. To see whether this contribution can be canceled by the contact term contribution, we can rewrite  the $\xi_2$ term in \eqref{contactg2} using the relation
\begin{equation}
    \begin{aligned}
         (\bar{v}_2^ \mu  u_1^\nu) (\bar{v}_{3}^\alpha \gamma_{\mu\nu} u_{4\al})=&~(\bar{v}_2^\mu u_{1\mu})(\bar{v}_3^\nu u_{4\nu})-(\bar{v}_2 ^\mu u_1^\nu )(\bar{v}_3^\al \gamma_{\nu}\gamma_\mu u_{4\al})\,,\\=& -\frac{3}{4}(\bar{v}_3^\mu u_{1}^\nu)(\bar{v}_{2\nu} u_{4\mu})- \frac{3}{4}(\bar{v}_3^\mu \gamma_5 u_1^\nu)(\bar{v}_{2\nu}\gamma_5 u_{4\mu})- \frac{1}{4}(\bar{v}_3^\mu \gamma_5 \gamma^\alpha u_1^\nu)(\bar{v}_{2\nu}\gamma_5 \gamma_\alpha u_{4\mu})\\&-\frac{3}{4}(\bar{v}_3^\mu \gamma^\al u_1^\nu)(\bar{v}_{2\nu}\gamma_\al u_{4\mu})+(\bar{v}_3^\al \gamma^\mu u_1^\nu)(\bar{v}_{2\mu}\gamma_\nu u_{4\al}) -\frac{1}{8}(\bar{v}_3^\mu \gamma^{\al\beta} u_1^\nu)(\bar{v}_{2\nu}\gamma_{\al\beta} u_{4\mu})\,.
    \end{aligned}
\end{equation}
In the second line, we have used the Fierz identity and $\gamma$-trace constraint. A similar equation can be obtained for $(\bar{v}_2 ^ \al  \gamma_{\mu\nu}u_{1\al}) (\bar{v}_{3}^\mu  u_{4}^\nu)$ by interchanging $(2\leftrightarrow 3)$ and $(1\leftrightarrow 4)$. The Fierz identities convert the $s$-channel amplitudes into a $t$-channel form, and this can be applied to the $\xi_1$ contact terms and the pole term \eqref{propwi}, as well. The sum of all contributions will involve only different contractions of $( 3   1  )( 2   4 )$ and no structure like $( 2   1  )( 3  4 )$ appears. 

In the total $\tilde{\cal A}_4$ expression written in such a form, the different spinor structures do not cancel, and therefore the Ward identity cannot be satisfied for any choice of the coefficients $l_1$, $l_2$. Alternatively, using the center-of-mass values in Appendix~\ref{app:explicitkin}, one can also explicitly check that $\tilde{\cal A}_4$ does not vanish.  In other words, single photon exchange is not compatible with the Ward identity.

\subsection{Adding a spin-2 graviton}
\label{sec:DiracSpin1Spin2}
 
Let us summarize the result obtained in Section~\ref{sec41}. Under the following assumptions: (i) the particle content is a photon and a charged Dirac spin-3/2 fermion $(\psi_\mu ,\chi_\mu)$, (ii) the four-point spin-3/2 scattering amplitude satisfies the Ward identity, and (iii) the dimension of the operators is at most 5,
the amplitude is constructible under the ALT shift, and we obtained two possible contact terms $\xi_1$, $\xi_2$ in \eqref{contactg1} and \eqref{contactg2}, respectively. However, this result contradicts assumption (ii), for any non-trivial choice of the coefficients $l_1$, $l_2$, $\xi_1$, $\xi_2$. Therefore, the above three assumptions are incompatible. If we want to keep assumption (ii), as we required in Section~\ref{sec3}, the price to pay is to modify assumptions (i) and/or (iii). Namely, either we introduce other degrees of freedom, or higher-derivative couplings. In the following, we investigate the first possibility.

What new particle can be introduced? Obviously, adding another photon amounts to a rescaling of the result and does not solve the problem. We can also consider massive spin-0 and spin-1, but as we showed in Section~\ref{sec3}, the four-point amplitudes constructed from these interactions do not have contact terms.
Meanwhile, for the Ward identity to be restored, we need  some new contact term contribution to cancel the one from massless spin-1 exchange, so the spin-0 and spin-1 cannot restore the Ward identity, either. 

The next simplest possibility is to introduce the spin-2 graviton.
Contact terms do arise from graviton exchange \cite{Gherghetta:2024tob} and may be able to cancel the photon contribution. When there is a single  Majorana spin-3/2 fermion,  the scattering of identical particles $\psi_\mu\psi_\mu\rightarrow \psi_\mu\psi_\mu$ involves three factorization channels ($s,t,u$). When these channels are summed, the Ward identity of this amplitude is satisfied.  If  there are two Majorana spin-$3/2$ fermions $(\chi_\mu,\psi_\mu)$, the scattering $\psi_\mu\chi_\mu\rightarrow\psi_\mu\chi_\mu$ has only one channel ($u$) with graviton coupling, shown in the diagram below:
\begin{figure}[ht]\hspace{4cm}
      \hspace{0.15\textwidth} 
  \begin{minipage}[t]{0.2\textwidth}
    \centering
\hspace{-1pt}\includegraphics[width=0.8\linewidth]{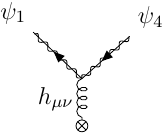} \\[1pt]  \includegraphics[width=0.8\linewidth]{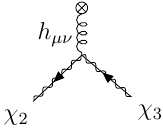}
  \end{minipage} 
\end{figure}

One can check that the Ward identity for this single channel is not satisfied. In this sense, if we had started with a Dirac spin-3/2 fermion $(\chi_\mu,\psi_\mu)$ and a graviton, without the photon, we would have met the same issue and again, the Ward identity would have required introducing new particles. Thus, our question is  whether the Ward identity can be recovered with both graviton and photon interactions. 

The three-point amplitude involving a graviton is given in \eqref{Vhpsipsiexp}. The $u$-factorization channel yields the following contact terms 
\begin{equation}
\frac{c_h^2 }{8} \kappa^2(\bar{v}_{2\al}\gamma_{\nu} u_{3\mu}-\bar{v}_{2\al}\gamma_{\mu} u_{3\nu}+\bar{v}_{2\nu}\gamma_{\mu} u_{3\al}-\bar{v}_{2\mu}\gamma_{\nu} u_{3\al}+\bar{v}_{2\nu}\gamma_{\al} u_{3\mu}-\bar{v}_{2\mu}\gamma_{\al} u_{3\nu})\bar{v}_{1}^{\mu}\gamma^{\al} u_{4}^{\nu}\,.
\end{equation}
 To check the Ward identity,   we replace $\bar{v}_2^\al$ by $\bar{v}_2 k_2^\alpha $ and take the massless limit. The contact term contribution to  $\tilde{\cal A}_4$ is straightforward, whereas the propagator pole part simplifies to
 \begin{equation}
     \tilde{\cal A}_4\supset -\frac{c_h^2}{2}\kappa^2\left(\bar{v}_2 \gamma_\mu u_{3\nu}+\bar{v}_2 \gamma_\nu u_{3\mu}\right)\left[\frac{1}{2}\bar{v}_1^\alpha\gamma^\mu(k_4^\nu -k_1^\nu)u_{4\alpha}-\bar{v}_1^\al k_{4\al}\gamma^\mu u_{4}^\nu +\bar{v}_1^\nu \gamma^\mu k_{1\alpha}u_4^\alpha \right]\,,
 \end{equation}
where the denominator $1/(k_1\cdot k_4)$ is automatically canceled. Consequently, in the total Ward identity, for the second and third lines in \eqref{propwi} to vanish, we still have to require $l_1+4l_2=0$. 
 
We  now proceed to check the Ward identity by summing up the propagator pole terms and contact terms with both photon and graviton exchanges. 
To simplify the calculation, we will evaluate the Ward identity in the center-of-mass frame using explicit kinematics. We first choose the helicities of $\psi_1,\chi_2,\chi_3,\psi_4$ to be $\left(+\frac{3}{2},+\frac{3}{2},+\frac{3}{2},+\frac{3}{2}\right)$, and then require
\begin{equation}
    \tilde{\cal A}_4^{\left(+\frac{3}{2},+\frac{3}{2},+\frac{3}{2},+\frac{3}{2}\right)}
    =3 \sqrt{2} E^3\left(-8\frac{l_2^2}{M^2}
    +c_h^2\kappa^2\right) \sin  \theta =0\,. 
\end{equation}
This results in two solutions for $l_2$.  Similarly, we can proceed for all the 16 helicity configurations $\left(\pm\frac{3}{2},\pm\frac{3}{2},\pm\frac{3}{2},\pm\frac{3}{2}\right)$. In the end, we find that the Ward identity can be satisfied with the following solutions
\begin{equation}
      l_1= \pm\sqrt{2}c_h M \kappa,\quad l_2=\mp\frac{ c_h M \kappa}{2\sqrt{2}},\quad \xi_1=c_h^2\kappa^2 ,\quad \xi_2=-\frac{c_h^2\kappa^2}{4}\,.\label{sec4sol}
\end{equation} 
These two solutions are equivalent up to the rescalings $A_\mu \rightarrow  -A_\mu$ and $e\rightarrow -e$. 
When $c_h=\frac{1}{2}$, they correspond exactly to the on-shell interactions in the minimal $N=2$ Lagrangian (see \cite{Ferrara:1976fu,Freedman:1976aw,Ferrara:1976iq}) in the limit $m_\psi\rightarrow0,m_\chi\rightarrow0$. 
The $\chi\chi\chi\chi$ and $\psi\psi\psi\psi$ contact terms can be reconstructed by computing  the $\chi_\mu\chi_\mu\rightarrow\chi_\mu\chi_\mu$ and $\psi_\mu\psi_\mu\rightarrow\psi_\mu\psi_\mu$
amplitudes using the ALT shift, which was already performed in \cite{Gherghetta:2024tob}, and the result also agrees with \cite{Ferrara:1976fu,Freedman:1976aw,Ferrara:1976iq}. In addition, by requiring a smooth massless limit for the photon coupling, we obtain the charge-mass relation \eqref{minimalcond}, and using \eqref{sec4sol} with $M=M_P$ and $c_h=1/2$, we obtain $m_{3/2}=\sqrt{2}eM_P$. This relation is consistent with gauged $N=2$ supergravity, where local supersymmetry relates the Dirac gravitino mass to the coupling and cosmological constant \cite{Deser:1977uq,Freedman:1976aw}. By tuning the cosmological constant to zero, supersymmetry is \textit{softly} broken and the gravitino becomes massive in Minkowski spacetime \cite{Deser:1977uq}.  The Lagrangian counterpart is \eqref{n2lagrangian} plus gravitino mass terms. Supersymmetry is restored by taking the massless limit  $m_{3/2}\rightarrow0$, and we recover the minimal $N=2$ supergravity Lagrangian.

\subsection{The problem with unitarity} 

The theory constructed so far in Section~\ref{sec:DiracSpin1Spin2} contains a massless photon, a graviton and a massive Dirac spin-$3/2$ fermion. The construction is self-consistent as it satisfies the three requirements at the beginning of Section~\ref{sec3}. We now investigate the UV behavior of the four-point scattering amplitude $\psi_\mu\chi_\mu \rightarrow\psi_\mu\chi_\mu$.

When evaluated in the center-of-mass frame,  we find that  the leading $E^6$ and $E^5$ terms are canceled by the sum of graviton and photon exchanges.  With $c_h=1/2$, the total amplitudes are
\begin{equation}
\begin{aligned}
&{\cal A}_{h+A }^{\left(\frac{1}{2},\frac{1}{2},\frac{1}{2},\frac{1}{2}\right)}
=-\frac{2}{3}(1+ \cos \theta)\,\frac{\kappa^2 E^4}{m_{3/2}^2}+\mathcal{O}(\kappa^2E^2)\,,\\
&{\cal A}_{h+A }^{\left(\frac{1}{2},\frac{1}{2},\frac{1}{2},-\frac{1}{2}\right)}
=\frac{2}{3}\sin\theta\,\frac{\kappa^2E^3}{m_{3/2} }+\mathcal{O}(\kappa^2E)\,.
\end{aligned}
\label{hAamp}
\end{equation}
The $E^4$ and $E^3$ terms are non-vanishing, so the theory is not unitary up to the Planck scale, which is expected because supersymmetry is not spontaneously broken. As considered in Section~\ref{sec3}, one may attempt to increase the unitarity cutoff obtained from \eqref{hAamp} by introducing scalar interactions, mimicking an ``$N=2$ Polonyi model''. Assuming that the scalar and pseudoscalar states couple to identical fermions as in \eqref{PScouplings}, with couplings $\{c_{S,\psi},c_{P,\psi}\}$ $\{c_{S,\chi},c_{P,\chi}\}$ for $\psi_\mu $ and $\chi_\mu$, respectively, we then have only the $u$-factorization channel, with
\begin{equation}
\begin{aligned}
&{\cal A}_{S+P }^{\left(\frac{1}{2},\frac{1}{2},\frac{1}{2},\frac{1}{2}\right)}=\mathcal{O}(\kappa^2E^2)\,,\\
&{\cal A}_{S+P }^{\left(\frac{1}{2},\frac{1}{2},\frac{1}{2},-\frac{1}{2}\right)}
=\frac{8}{9}c_{S,\psi}c_{S,\chi}
\sin  \theta (1-\cos\theta) 
\frac{ \kappa^2E^3 }{m_{3/2} }+\mathcal{O}(\kappa^2E)\,.
\end{aligned}
\label{PSamp}
\end{equation}
Clearly, \eqref{PSamp} is insufficient to cancel \eqref{hAamp}, and therefore we need to add more particle states.

We next introduce a massive spin-$1$ $B_\mu$ that is also assumed to couple to identical fermions as in \eqref{threeptB}, with couplings $ c_{B,\psi} $ for $\psi_\mu $ and $ c_{B,\chi} $  for $\chi_\mu$. The $B_\mu$ exchange amplitudes are
\begin{equation}
\begin{aligned}
&{\cal A}_{B}^{\left(\frac{1}{2},\frac{1}{2},\frac{1}{2},\frac{1}{2}\right)}
= \frac{16}{9}c_{B,\psi}c_{B,\chi}(1-\cos\theta)\frac{E^4}{m_{3/2}^4}+  \mathcal{O}(E^2/m^2)\,,\\
&{\cal A}_{B}^{\left(\frac{1}{2},\frac{1}{2},\frac{1}{2},-\frac{1}{2}\right)}
= 8c_{B,\psi}c_{B,\chi}
\sin\theta (1-\cos\theta) 
\frac{E^3}{m_{3/2}^3}+\mathcal{O}(E/m)\,.
\end{aligned}
\label{Bamp}
\end{equation}
In the combination of \eqref{hAamp}, \eqref{PSamp}, \eqref{Bamp}, the leading $E^4$, $E^3$ energy behavior is still not canceled. We may try to generalize the scalar interactions by adding a mixed term
\begin{equation}
    \begin{aligned} c_S^\prime  \kappa m_{3/2} \psib^\mu\chi_\mu S+i c_P^\prime \kappa  \varepsilon^{\mu\nu\alpha\beta } \partial_\mu P \psib_\nu \gamma_\alpha\chi_\beta \,,
     \end{aligned} 
\end{equation}
which creates $s$- and $t$-factorization channels, so that in addition to \eqref{PSamp}, we also have
\begin{equation}
\begin{aligned}
&{\cal A}_{S+P,st }^{\left(\frac{1}{2},\frac{1}{2},\frac{1}{2},\frac{1}{2}\right)}
=-\frac{16}{9}(c_S^{\prime 2}+4c_P^{\prime 2})\frac{\kappa^2E^4}{m_{3/2}^2}+ \mathcal{O}(\kappa^2E^2)\,,\\
&{\cal A}_{S+P ,st}^{\left(\frac{1}{2},\frac{1}{2},\frac{1}{2},-\frac{1}{2}\right)}
=\frac{8}{9}c_S^{\prime 2}
\sin\theta (1+\cos\theta) 
\frac{\kappa^2E^3}{m_{3/2}}+\mathcal{O}(\kappa ^2 E/m)\,.
\end{aligned}
\label{PSamp1}
\end{equation}

There is no solution, either, which is consistent with known results in supergravity. More precisely, one cannot break $N=2$ supergravity directly to $N=0$ and generate a Dirac gravitino mass term, using hypermultiplets only. In general, the spontaneous breaking of extended supergravity simultaneously occurs with the spontaneous breaking of the gauged isometries, with spin-$1$ gauge bosons. In this case, a mass for the spin-$1$ boson is generated after the breaking \cite{Andrianopoli:2002rm}. As will be shown in Sections~\ref{partialsec} and \ref{sec5}, the massive spin-$1$ gauge boson plays an essential role in the broken theory.

\subsection{Partial breaking of $N=2$ supergravity}\label{partialsec}
 
A well-known mechanism to spontaneously break $N=2$ supergravity is  the    partial  breaking to $N=1$ supergravity\cite{Ferrara:1995xi}. This process requires a hypermultiplet (with two Weyl fermions and two complex scalars), and a gauge multiplet (with one spin-1, two spin-$1/2$, and a complex scalar), in addition to the gravity multiplet (with one spin-2, two spin-$3/2$, and a spin-1). After the breaking, the two spin-$1$ bosons become massive. With a certain choice of parameters, one of the spin-3/2 fermions remains massless, and a massive $N=1$ multiplet with spins $\left(\frac{3}{2},1,1,\frac{1}{2}\right)$ appears.

The three-point amplitudes relevant for the partial breaking involve one massive spin-$3/2$ fermion ($\psi_\mu$) and one massless spin-$3/2$  fermion ($\chi_\mu$). The identical fermions can still couple to the graviton via \eqref{mincouponshell}. When the fermion is massless, one obtains the three-point amplitude by taking the massless limit of \eqref{mincouponshell}. Similar to what we have discussed for the Dirac gravitino, the Ward identity for $\psi \chi\rightarrow \psi\chi $ cannot be satisfied with graviton exchange only, and the spin-1 is necessary to complement the diagrams with $s$- and $t$-channels. The three-point amplitudes with $m_\psi\neq0,m_\chi=0$ are given in \eqref{Apsichi}-\eqref{equalBpsi}, and we can further impose the requirements that there are no higher derivatives and the massless limit is smooth.  The massless spin-$1$ amplitudes \eqref{Apsichi} are excluded because they originate from higher-derivative couplings. For the massive spin-$1$ coupling with $m_B\neq m_\psi$, the $\tilde{b}_1^\pm$ amplitudes have higher derivatives, whereas the $\tilde{b}^\pm_2$ amplitudes can be obtained by one-derivative couplings, but are divergent in the  massless limit. For example, in the massless limit, the $\left(-\frac{3}{2},0,-\frac{3}{2}\right)$ helicity amplitude becomes
\begin{equation}
\frac{1}{\sqrt{2}}\tilde{b}^-_2 m_\psi \left<13\right>^3\,,
\end{equation}
where $\tilde{b}_2^-$ has mass dimension $-3$. Thus, even if we assume that $\tilde{b}_2^-\propto \kappa/m^2$, with just one factor of $\kappa$, the above amplitude behaves as $1/m$, which is singular.

Finally, we are left with the equal mass $m_B=m_\psi$ case. In Eq.~\eqref{equalBpsi}, we can check that this form of the amplitude can originate from $\leq 1$ derivative couplings, because $3$ appears as $\varepsilon_3\left|3\right>$ or $\varepsilon_3\left|3\right]$, $\btwo$  appears as $\left|\btwo\right>\left|\btwo\right>\left|\btwo\right]$ or $\left|\btwo\right>\left|\btwo\right]\left|\btwo\right]$. The only derivative can act on the spin-1 when it appears as $\left|\bone\right>\left|\bone\right>$ or $\left|\bone\right]\left|\bone\right]$. In addition, the massless limit of \eqref{equalBpsi} gives:
\begin{equation}   \begin{aligned} &{\cal A}_B^{(+\frac{3}{2})}\rightarrow
    \left \{\begin{aligned}
         &\sqrt{2}\tilde{b}^+_3 m_B \frac{[23]^4 }{[12][13]}\quad \text{for} \quad \left(-1,\frac{3}{2},\frac{3}{2}\right)\\&\frac{2}{\sqrt{6}}\tilde{b}_3^+ m_B\frac{[13]^3 }{[12]}
    \quad \text{for} \quad \left(1,-\frac{1}{2},\frac{3}{2}\right) \\& -\frac{2}{\sqrt{3}}\tilde{b}^+_3m_B\frac{[13][23]^2}{[12]}
    \quad \text{for} \quad \left(0,\frac{1}{2},\frac{3}{2}\right) \end{aligned}\right.\\&{\cal A}_B^{(-\frac{3}{2})}\rightarrow
    \left \{\begin{aligned}
         & \sqrt{2}\tilde{b}^-_3m_B\frac{\left<23\right>^4 }{\left<12\right>\left<13\right>}
    \quad \text{for} \quad \left(1,-\frac{3}{2},-\frac{3}{2}\right)\\&  \frac{2}{\sqrt{6}}\tilde{b}^-_3 m_B\frac{\left<13\right>^3 }{\left<12\right>}
    \quad \text{for} \quad \left(-1, \frac{1}{2},-\frac{3}{2}\right)\\&-
    \frac{2}{\sqrt{3}}\tilde{b}^-_3  m_B \frac{\left<13\right>\left<23\right>^2 }{\left<12\right>}\quad \text{for} \quad \left(0,-\frac{1}{2},-\frac{3}{2}\right)\end{aligned}\right.
\end{aligned}
\label{masslessB1}
\end{equation} 
where $\tilde{b}_3^\pm$ has mass dimension $-2$. Hence, as long as $\tilde{b}_3^\pm\propto \kappa/m_B$, the massless limit is smooth.

In summary, imposing the Ward identity, no higher derivatives and a smooth massless limit, we find that in addition to the gravitational coupling,  the massless spin-$3/2$ fermion $\chi_\mu$, and the massive spin-$3/2$ fermion, $\psi_\mu$ must couple to the massive spin-$1$ boson $B_\mu$, with $m_B=m_\psi$. This result, derived from the bottom-up, is consistent with the partial breaking mechanism \cite{Ferrara:1995xi}.

More general solutions can be obtained by considering both spin-$3/2$ fermions to be massive and then the massive photon does not necessarily have the same mass as the spin-$3/2$ fermion.  In the next section, we will consider the case with two massive Majorana spin-$3/2$ fermions that first couple to a massive spin-$1$ boson. We will show that unitarity is possible with different solutions and additional particle content.

\section{$N=2$ supergravity from a massive spin 1}\label{sec5}

Previously, we have investigated two Majorana spin-$3/2$ fermions coupling to a massless photon, and shown that the scattering amplitude grows faster than $E^2$ at high energy, regardless of the additional particle content.  
To search for a theory with at most $E^2$ growth, thereby realizing a unitarity cutoff that can be independent of particle masses,
we now start by coupling the Majorana spin-3/2 fermions to a \textit{massive} spin-$1$ boson.

\subsection{Spin-$3/2$ coupled to a massive photon}

The most general three-point amplitudes involving a massive photon and two massive spin-$3/2$ fermions are derived in \eqref{generalmassive1}, which reduce to 10 independent structures using the Schouten identities \eqref{schouten1}-\eqref{schouten2}. We now consider a massive spin-$1$ boson coupling to two different fermions $\psi_\mu$, $\chi_\mu$, therefore, Fermi statistics no longer apply. To remain general, we assume the fermion masses $m_{\psi,\chi}$ can be unequal and label the spin-1 particle by $\bone$, and $\psi_\mu$, $\chi_\mu$ by $\btwo$, $\bthree$, respectively.

We first discard the amplitudes $[\btwo\bthree]^2[\bone\btwo][\bone\bthree]$, $\left<\btwo\bthree\right>^2\left<\bone\btwo\right>\left<\bone\bthree\right>$ because they originate from higher-dimensional operators. As we have checked in \eqref{masslessB}, 
$\left<\btwo\bthree\right>\left[\btwo\bthree\right] \left<\bone\btwo\right>\left[\bone\bthree\right] $  and   $\left<\btwo\bthree\right>\left[\btwo\bthree\right] \left<\bone\bthree\right>\left[\bone\btwo\right]\  $ have a smooth massless limit as long as the coupling scales as $\kappa/m^2$,\footnote{In \eqref{masslessB}, the two amplitudes appear with a relative minus sign, but the smooth massless limit also holds for arbitrary relative coefficients.}  where  $m$ is a particle mass. The remaining six amplitudes do not independently have a smooth massless limit, but combinations of them could. We can write
\begin{equation}
    \begin{aligned}
{\cal A}_B=&y_1\left<\btwo\bthree\right>[\btwo\bthree]\left<\bone\btwo\right>\left<\bone\bthree\right>+
y_2\left<\btwo\bthree\right>[\btwo\bthree]\left[\bone\btwo\right]\left[\bone\bthree\right]+y_3[\btwo\bthree]^2\left<\bone\btwo\right>\left[\bone\bthree\right]\\&+y_4[\btwo\bthree]^2\left[\bone\btwo\right]\left<\bone\bthree\right>+y_5    \left<\btwo\bthree\right>^2 \left<\bone\btwo\right>\left[\bone\bthree\right]+y_6    \left<\btwo\bthree\right>^2 \left[\bone\btwo\right]\left<\bone\bthree\right>\,\\&+y_7\left<\btwo\bthree\right>\left[\btwo\bthree\right] \left<\bone\btwo\right>\left[\bone\bthree\right] +y_8\left<\btwo\bthree\right>\left[\btwo\bthree\right] \left<\bone\bthree\right>\left[\bone\btwo\right]\   ,
\end{aligned}
\end{equation}
where $y_i$ are arbitrary coefficients with mass dimension $-3$.
Denoting for now the spin-1 mass by $m_1$, the dominant amplitudes in the 
high-energy limit ($E\gg m_{1,\psi,\chi}$) are
\begin{equation}
    {\cal A}_B\rightarrow \left\{\begin{aligned}&
        \frac{1}{\sqrt{3}} (y_1 m_\psi-y_5 m_1) \left< 12 \right>^2\left< 23 \right>
        \\&\frac{1}{\sqrt{3}}(  y_6 m_1 -  y_1 m_\chi)\left< 13 \right>^2\left< 23 \right>
        \\& \frac{1}{\sqrt{3}}( y_3 m_1 -  y_2 m_\chi)\left[  13 \right]^2\left[  23 \right]
        \\& \frac{1}{\sqrt{3}}(  y_2m_\psi -  y_4m_1)\left[  12 \right]^2\left[  23 \right]
        \\& \frac{1}{\sqrt{2}}(y_4 m_\chi -y_3 m_\psi)[23]^3
        \\& \frac{1}{\sqrt{2}}(y_5 m_\chi -y_6 m_\psi)\left<23\right>^3
    \end{aligned} \right. 
    \end{equation}
By dimensional analysis, the coupling coefficients $y_i\propto 1/m^3$, and thus diverge in the massless limit ($m\rightarrow 0$). However, a smooth massless limit exists if 
\begin{equation}
        \begin{aligned}
            y_3=\frac{m_\chi}{m_1} y_2,\quad 
            y_4=\frac{m_\psi}{m_1} y_2,\quad 
            y_5=\frac{m_\psi}{m_1} y_1,\quad 
            y_6=\frac{m_\chi}{m_1} y_1\,,
        \end{aligned}
        \label{smoothym}
    \end{equation}
where the next order $(\mathcal{O}(y_i m^2))$ is not divergent provided $y_i\propto \kappa/m^2$.
There are two types of interactions satisfying the conditions \eqref{smoothym}. We introduce   massive vector fields $B_\mu$ and $B_\mu^\prime $  for the CP-even and CP-odd interactions, respectively.
Their field strengths are given by $G_{\mu\nu}=\partial_\mu B_\nu -\partial_\nu B_\mu$, 
$G^\prime _{\mu\nu}=\partial_\mu B_\nu^\prime  -\partial_\nu B_\mu^\prime $. The spin-$1$ mass $m_1$ can be either $m_B$ or $m_{B^\prime}$.
We then have the on-shell interactions
\begin{equation}
   i \alpha_0\bar{\chi}_\mu \slashed{B}\psi^\mu +\frac{\alpha_1}{M} \bar{\chi}_\mu G^{\mu\nu}\psi_\nu +\frac{\alpha_2}{M} \bar{\chi}_{\mu } G^{\al\beta}\gamma_{\al\beta}\psi^\mu\,,
   \label{Gint}
\end{equation} 
\begin{equation}
    \beta_0\bar{\chi}_\mu \gamma_5\slashed{B}^\prime\psi^\mu +i\frac{\beta_1}{M}\bar{\chi}_\mu \gamma_5 G^{\prime\mu\nu}\psi_\nu +i\frac{\beta_2}{M}\bar{\chi}_{\mu }  \gamma_5G^{\prime\al\beta}\gamma_{\al\beta}\psi^\mu\,,
   \label{Gtildeint}
   \end{equation}  
where $M$ is a generic mass scale and imposing \eqref{smoothym} requires $\alpha_1+4\alpha_2=0$, $\beta_1+4\beta_2=0$. Note also that $\alpha_0$, $\beta_0$ must vanish in the massless limit because the associated three-point amplitudes are proportional to $\alpha_0/m$ or $\beta_0/m$ in this limit. In the $N=2$ supergravity Lagrangian coupled to a vector multiplet, it has been noticed \cite{Zinovev:1990mw} that the two vector fields in the gravity multiplet (``graviphoton'') and in the vector multiplet appear as $(B_\mu\pm i\gamma_5B_\mu^\prime )$, and on-shell, their interactions with the gravitino are of the form  \eqref{Gint}-\eqref{Gtildeint} (when the scalar field obtains a VEV). The interactions \eqref{Gint} and \eqref{Gtildeint} will be our starting point to reconstruct supergravity.

Similar to the method used in Section~\ref{sec4}, we employ the ALT shift to compute the $\psi_1\psi_2\rightarrow\chi_3\chi_4$ scattering amplitude.\footnote{The $\psi_1\psi_2\rightarrow\chi_3\chi_4$ amplitude is related to $\psi_1\chi_2\rightarrow\psi_3\chi_4$ by crossing symmetry, but the former is more convenient in the case of unequal masses, because there is a single energy scale in the center-of-mass frame.} There are two $(t,u)$ factorization channels associated with each massive spin-1.  By $B_\mu$ exchange, we recover   the same $\xi_1$, $\xi_2$ contact terms as in \eqref{contactg1}-\eqref{contactg2}, with $\xi_1+2\xi_2= (\alpha_1+2\alpha_2)^2/M^2$, and we relabel the polarization vectors by  $(1\leftrightarrow3)$.
The $B_\mu^\prime$ exchange gives rise to the new contact terms 
\begin{equation}  
\xi^\prime_1 \left[(\bar{v}_2^ \mu \gamma^5 u_4^\nu)(\bar{v}_{1\mu} \gamma^5u_{3\nu})-(\bar{v}_2^\mu \gamma^5u_{4 \nu})(\bar{v}_1^\nu \gamma^5u_{3\mu})\right]-\left( 1\leftrightarrow 2\right)\,, 
\end{equation}
and
\begin{equation} 
\xi_2^\prime\left[ (\bar{v}_2^ \mu \gamma^5 u_4^\nu) (\bar{v}_{1}^\alpha \gamma^5\gamma_{\mu\nu} u_{3\al})+(\bar{v}_2^ \al  \gamma^5\gamma_{\mu\nu}u_{4\al}) (\bar{v}_{1}^\mu\gamma^5  u_{3}^\nu)\right] -\left( 1\leftrightarrow 2\right)\,,
\end{equation}
with $\xi^\prime_1+2\xi^\prime_2= -(\beta_1+2\beta_2)^2/M^2 $. 

For the Ward identity, we encounter the same situation as in Section~\ref{sec4}. With only the $(t,u)$ channels, it is impossible to satisfy the Ward identity, and the graviton is again necessary to recover it. Because the graviton couples to identical fermions, it supplements the $\psi_1\psi_2\rightarrow\chi_3\chi_4$ scattering with an $s$-factorization channel. In addition, the mass scale $M=1/\kappa$ is now related to the Planck scale. The Ward identity is then realized for
\begin{equation}
\begin{aligned}
  &  \alpha_2= 
\pm\sqrt{\frac{c_h^2}{8}-\beta_2^2}
  =-\frac{1}{4}\alpha_1
,\quad \xi_2= -\xi_2^\prime -\frac{1}{4}(c_h^2-16 \beta_2^2)\kappa^2\,,\\
&\xi_1=2\xi_2^\prime +(c_h^2 -12\beta_2^2)\kappa^2 ,\quad \xi_1^\prime=-2\xi_2^\prime -4\beta_2^2\kappa^2\,.
\end{aligned}\label{genwardid}
    \end{equation}
Note that since $c_h\neq 0$, the couplings $\alpha_1,\alpha_2\neq0$, so the coupling to $B_\mu$ is unavoidable, whereas $\beta_1,\beta_2$ could be vanishing according to \eqref{genwardid}.

Let us comment on a special limit where one of the spin-$3/2$ particles is massless ($m_\chi\rightarrow0$), i.e.~$N=2$ broken to $N=1$ supergravity. We have shown in Section~\ref{partialsec} from the smooth massless limit of the three-point amplitudes   that the spin-$3/2$ and spin-$1$ masses are equal (i.e. $m_\psi=m_B$), if $m_\chi=0$.
For the four-point amplitude, since the Ward identity  is  imposed when all the particles are massless, the same constraints \eqref{genwardid} also apply to the theory with $m_\chi\rightarrow0$. The high-energy behavior of the $\psi\psi\rightarrow \chi\chi $ amplitude  is dominated by $\mathcal{O}(\kappa^2 E^4/m^2)$, when $\psi_\mu$ is longitudinal and $\chi_\mu$ can only be transverse corresponding to a massless state. Due to the Ward identity, this dominant behavior is canceled between the propagator pole and the contact terms, leaving the next order $\mathcal{O}(\kappa^2 E^2)$ which implies a Planck scale cutoff. Next, for the $\psi\psi\rightarrow \psi\psi $ amplitude to be unitary, the same discussion for $N=1$ in Section~\ref{sec3} applies. Therefore, in addition to the graviton and the interactions \eqref{Gint}-\eqref{Gtildeint}, one has to introduce the usual $F$-term (or $F$- plus $D$-term) supersymmetry breaking interactions, satisfying \eqref{conditionmA}. To derive the complete mass spectrum and match to the partial-breaking mechanism \cite{Ferrara:1995xi}, other scattering amplitudes such as $BB\rightarrow\psi\psi$ can be studied as well.

\subsection{Unitarity solutions to ${\cal O}(\kappa^2 E^2)$}

Without the massive spin-1, the unitarity cutoff $\sqrt{ m_{3/2}/\kappa}$ depends on the particle mass.
In order to obtain a unitarity cutoff independent of the particle mass we now search for a solution so that the spin-3/2 scattering has a Planck scale cutoff by including the massive spin-1 contribution. First of all, although we are studying an $N=2$ theory (with two Majorana spin-3/2 fermions), the constraints from $N=1$ must also be satisfied. Namely, the amplitude of identical fermion scattering $\psi_1\psi_2\rightarrow\psi_3\psi_4$ or $\chi_1\chi_2\rightarrow\chi_3\chi_4$ should not grow faster than $E^2$ at high energy. As we have discussed in detail in Section~\ref{sec3}, this requires additional scalar, pseudoscalar, and massive spin-1 interactions. When the massive spin-1 boson couples to identical fermions, the on-shell interaction is CP-odd (see Eq.~\eqref{threeptB}), and therefore, we associate this spin-$1$ state with $B^\prime_\mu$:
\begin{equation}
    c_{B^\prime,\psi}  \bar{\psi}^\mu \slashed{B}^\prime \gamma^5\psi_\mu\,+
    c_{B^\prime,\chi}  \bar{\chi}^\mu \slashed{B}^\prime \gamma^5\chi_\mu\,.
  \label{threeptBprime}
\end{equation}
Likewise, we generalize the scalar couplings   \eqref{PScouplings} to
\begin{equation}
    \begin{aligned} 
    c_{S,\psi} \kappa m_{\psi}  \psib^\mu\psi_\mu S+i c_{P,\psi}\kappa  \,\varepsilon^{\mu\nu\alpha\beta } \partial_\mu P \psib_\nu \gamma_\alpha\psi_\beta \,+c_{S,\chi} \kappa m_{\chi}  \bar{\chi} ^\mu\chi_\mu S+i c_{P,\chi}\kappa  \,\varepsilon^{\mu\nu\alpha\beta } \partial_\mu P \bar{\chi}_\nu \gamma_\alpha\chi_\beta \,.
\end{aligned}
\label{PScoupling4}
\end{equation}
The unitarity constraints were derived in Section~\ref{sec:Majoranaspin2} and therefore generalizing the result \eqref{conditionmA} for two fermions we obtain
\begin{equation}
 \begin{aligned}
        &c_{B',\chi (\psi)}^2= -\left(c_{S,\chi(\psi)}^2-\frac{3}{8}\right) \kappa^2 m_{\chi (\psi)}^2,\quad c_{P,\chi(\psi)}^2=\frac{1}{4}c_{S,\chi(\psi)}^2 -\frac{c_{B^\prime,\chi(\psi)}^2}{\kappa^2 m_{B'}^2}\,. 
 \end{aligned}\label{generalcs}
\end{equation}

Next we calculate the $\psi_1\psi_2\rightarrow\chi_3\chi_4$ and $\psi_1\chi_2\rightarrow\psi_3\chi_4$ scattering amplitudes by using the recursion relation (see \eqref{four-point-fact}), assuming all the on-shell interactions \eqref{mincouponshell}, \eqref{Gint}, \eqref{Gtildeint}, \eqref{threeptBprime} and \eqref{PScoupling4}. For general $m_\chi$ and $m_\psi$, we require that the high-energy behavior of both amplitudes 
is at most $\mathcal{O}(\kappa^2s)$, where $s=(E_1+E_2)^2$ is the center-of-mass energy. Without loss of generality, we will choose the sign convention $\alpha_2=\sqrt{\frac{1}{8}c_h^2-\beta_2^2}$ in \eqref{genwardid}. First of all, the cancellation of $\mathcal{O}(\kappa^2s^{5/2}/m^3)$ results in the mass relations
\begin{equation}
    m_\psi=
    \frac{16}{\kappa}\left(\alpha_0\sqrt{\frac{1}{32}-\beta_2^2}+\beta_0\beta_2\right),\quad      
    m_\chi 
=\frac{16}{\kappa}\left(\alpha_0\sqrt{\frac{1}{32}-\beta_2^2}-\beta_0\beta_2\right).
\end{equation}
We infer that $\alpha_0> 0$, otherwise   the masses would be negative. For the equal mass case ($\beta_0=0$ or $\beta_2=0$), there is no solution for  $\mathcal{O}(\kappa^2s^2/m^2)$ and $\mathcal{O}(\kappa^2s^{3/2}/m)$ terms to cancel, and therefore (i) the masses must be unequal: $m_\psi\neq m_\chi$, (ii) both massive spin-1, $B_\mu$ and $B^\prime_\mu$, are necessary. This is also the case in the  mechanism \cite{Ferrara:1995xi} where an $N=2$  vector multiplet participates in the breaking. In this case a mass splitting appears between the two spin-3/2 fermions, and the two spin-1 bosons in the vector and gravity multiplets obtain  masses via the Higgs mechanism. 

Imposing unitarity to $\mathcal{O}(\kappa^2s)$, we find a class of solutions with unequal masses. In particular, the simplest solution is
\begin{equation}
    \begin{aligned}
   &     m_\psi=\frac{2}{\kappa }(\alpha_0-\beta_0),\quad m_\chi=\frac{2}{\kappa }(\alpha_0+\beta_0),\quad m_B=\frac{-4\beta_0 }{\kappa },\quad m_{B'}=\frac{4\alpha _0}{\kappa},\\
  & c_{B,\psi}= 0,\quad c_{P,\psi } =\frac{1}{2}c_{S,\psi } ,\quad \beta_1=-4\beta_2=\frac{1}{2},\quad \alpha_1=-4\alpha_2=-\frac{1}{2},\\&c_{B,\chi}=\sqrt{\frac{3}{2}-\frac{1}{16 c_{S,\psi}^2}}(\alpha_0+ \beta_0),\quad c_{P,\chi}=\frac{3}{16 c_{S,\psi}},\quad c_{S,\chi}=-\frac{1}{8 c_{S,\psi}}\end{aligned}\label{specialsol}
\end{equation}
where $\alpha_0\geq-\beta_0>0$. Interestingly, the masses satisfy $m_B=m_\psi-m_\chi$, $m_{B'}=m_\psi+m_\chi$, which coincides with the relations in the mechanism  \cite{Ferrara:1995xi}, or more generally the spectrum of $N=2$ supergravity spontaneously broken to $N=0$  in the Minkowski vacuum.
This mass spectrum is continuously connected to that obtained from the partial breaking of $N=2$ supergravity to $N=1$   \cite{Zinovev:1990mw,Abe:2019svc}, since if we set $m_B=m_{B'}$, then one of the gravitinos is massless, implying that one of the supersymmetries is unbroken.

There are other more complicated solutions that lead to unitarity for spin-$3/2$ scattering. To derive the full constraints and the complete mass spectrum, other scattering amplitudes should be examined. For example, the $\psi\psi\rightarrow BB$ amplitude can a priori  be unitarized by scalar, spin 1, and graviton exchanges, 
which imposes constraints on different couplings to $B_\mu$. Similar techniques can be employed to derive these couplings on-shell and we leave this study for future exploration. 

\vspace{4mm}

Finally, let us summarize the IR and UV interactions that are involved in the (super)Higgs mechanism, in the $N=2$ case: 
\begin{center}
    \begin{tabular}{c|c}
IR (massive)   & UV  (massless)\\ \hline
 $\psi_\mu\psi_\nu h_{\alpha\beta}$, $\chi_\mu\chi_\nu h_{\alpha\beta}$     & $ \psi_{T\mu}\psi_{T\nu}  h_{\alpha\beta} $, $ \psi_{L}\psi_{L}  h_{\alpha\beta} $, $ \chi_{T\mu}\chi_{T\nu}  h_{\alpha\beta} $, $ \chi_{L}\chi_{L}  h_{\alpha\beta} $ \\  
 $\psi_\mu\psi_\nu\phi$, $\chi_\mu\chi_\nu\phi$     &   $ \psi_{T\mu }\psi_{L}  \phi_H $, $ \chi_{T\mu }\chi_{L}  \phi_H $  \\  
$\psi_\mu\psi_\nu B_\alpha$, 
 $\chi_\mu\chi_\nu B_\alpha$    & $ \psi_{T\mu }\psi_{L}  A_\alpha $, $ \psi_{T\mu }\psi_{L}  \phi_E $, $ \chi_{T\mu }\chi_{L}  A_\alpha $, $ \chi_{T\mu }\chi_{L}  \phi_E $\\
 
 $\psi_\mu\chi_\nu B_\alpha$    & $ \psi_{T\mu }\chi_{L}  A_\alpha $, $ \psi_{T\mu }\chi_{L}  \phi_E $, $ \chi_{T\mu }\psi_{L}  A_\alpha $, $ \chi_{T\mu }\psi_{L}  \phi_E $  \\
  $B_\mu B_\nu \phi$  &  $A_\mu \phi_H\phi_E$
\end{tabular}
\end{center}
The first two rows are the same as the minimal $N=1$ case: there are two separate superHiggs mechanisms for $\psi_\mu$ and $\chi_\mu$ and unitarity can be guaranteed for $\psi\psi\rightarrow\psi\psi$ and $\chi\chi\rightarrow\chi\chi$ scatterings. However, $\chi\chi\rightarrow\psi\psi$ does not have a Planck scale cutoff without additional interactions, because this amplitude has only an $s$-channel from graviton exchange.  
To include the $t$- and $u$-channels necessary for unitarity, $\psi_\mu$ and $\chi_\mu$ must interact directly. First, mixing the two spin-$3/2$ states via the scalar interaction ($\psi_\mu\chi_\nu \phi$) is forbidden,  
because in our on-shell setup, $\psi_\mu$, $\chi_\mu$ are \textit{physical} fermions. If they mix with $\phi=H-\left<H\right>$, the VEV $\left<H\right>$ then creates an off-diagonal term in the fermion mass matrix (or a kinetic mixing in the case of derivative couplings), which contradicts with the on-shell construction.  Alternatively, mixing the fermions with a massless spin-1 photon ($\psi_\mu\chi_\nu A_\alpha$) reduces the high-energy behavior of the four-point scattering to $\mathcal{O}(\kappa^2 E^4/m^2)$, but it does not restore unitarity as we have shown in Section~\ref{sec4}. It is then necessary to have an interaction, $\psi_\mu\chi_\nu B_\alpha$ with at least two massive spin-1, $B_\mu$ and $B_\mu^\prime$. The structure of the UV limit of this solution further proves that \textit{for $N=2$, superHiggs always arises with the Higgs mechanism}.

This conclusion also carries to $N>2$ (with more spin-$3/2$ particles), since the scattering of any pair of spin-$3/2$ $\chi\chi\rightarrow\psi\psi$ 
is still subject to the same discussion. In the meantime, just as the constraints \eqref{generalcs} from $N=1$  are imposed on $N=2$,  unitarity for $N>2$ will require the couplings to satisfy all the $N\leq 2$ constraints. For example, in $N=4$, in addition to identical fermion scatterings $\psi_i\psi_i\rightarrow\psi_i\psi_i$ that impose the conditions of the form \eqref{generalcs}, the scatterings of two fermions $\psi_i\psi_i\rightarrow\psi_j\psi_j$ ($i\neq j$) yield the constraints of $N=2$ type. On top of that, new scatterings $\psi_i\psi_j\rightarrow\psi_k\psi_l$ ($i\neq j\neq k\neq l$) further put constraints on unitarity solutions.

\section{Conclusion}

We have used on-shell methods to construct effective theories of massive spin-3/2 particles interacting with spin 0, 1 and 2 bosons. These methods rely on the ALT momentum shift~\cite{Ema:2024vww} which allows the constructibility of four-point scattering amplitudes for spin-3/2 particles. Starting with the primitive set of on-shell three-point amplitudes (containing up to one derivative), together with the assumption that the three-point amplitudes are smooth and satisfy the Ward identity in the massless limit ($m_{3/2}\rightarrow 0)$, we are able to construct four-point scattering amplitudes of massive spin-3/2 particles from the bottom up. These amplitudes are then used, together with imposing (perturbative) unitarity, to obtain the possible effective theories valid up to a cutoff $\Lambda\gg m_{3/2}$ that is independent of particle masses. 

In the case of one Majorana spin-3/2 particle, it is not possible to obtain an effective theory valid up to the UV scale $\Lambda$ (independent of the particle masses) when there are interactions only with spin 0 and 1 bosons. Interestingly, what is required is an interaction with a spin-2 graviton, that introduces an new dimensionful coupling $\kappa$, related to the Planck scale. In this case, effective theories valid up to $E\sim M_P$ are possible with complex scalars and massive photons. These effective theories, derived from the bottom-up,  correspond to well-known results in $N=1$ supergravity with $F$-and $D$- term breaking.

Other possibilities exist when there are {\it two} Majorana spin-3/2 fermions. Again, in this case, a gravitational interaction with a spin-2 graviton is required to obtain an effective theory valid up to $\Lambda\gg m_{3/2}$. However, in the special case of a Dirac spin-$3/2$ state charged under the U(1) symmetry corresponding to a massless photon, the cutoff $\Lambda$ still depends on the particle mass. This reproduces the well-known result that in a Minkowski background, supergravity is only softly-broken with a unitarity cutoff $\Lambda =\sqrt{m_{3/2} M_P}$ for a charged, Dirac gravitino.  
To obtain a Planck scale unitarity cutoff (independent of the particle masses) on the spin-$3/2$ scattering  amplitudes, we find solutions only for {\it unequal} Majorana spin-3/2 masses, where the spin-$3/2$ fermions necessarily couple to two massive photons that have both vector and axial-vector interactions.  
This possibility corresponds to $N=2$ supergravity spontaneously broken to $N=0$. The simplest solution (see \eqref{specialsol}) relates the two massive photon masses to the sum and difference of the gravitino masses, respectively, which coincides with known results in supergravity  \cite{Ferrara:1995xi,Zinovev:1990mw,Abe:2019svc}. Special cases where one of the spin-3/2 fermions is massless, also 
require a massive photon coupling which is consistent with the partial breaking mechanism of $N=2$ supergravity to $N=1$~\cite{Ferrara:1995xi}. To reproduce the complete mass spectrum and couplings, the same exercise can be repeated, by considering other scattering amplitudes (such as with massive spin-1) and requiring unitarity.

Our results, derived from the bottom up, can be extended to include and reconstruct the scalar potential in supergravity theories as well as the spin-1/2 (chiral fermion or gaugino interactions) by similarly imposing the same requirements, reproducing the complete supergravity structure. In this respect, it may also be useful to  supersymmetrize the ALT shift. Even though we only considered two-derivative supergravity interactions, it is also possible to use our bottom-up procedure to on-shell construct higher-derivative supergravity. While the naive dimensional analysis restricts the interactions to contain up to one derivative, the results in \cite{Ema:2025qgd} suggest that there is a wider class of constructible theories. In addition, applications to higher $N$ supergravity, supergravity in curved space (such as AdS), massive spin-2 states coupled to supergravity~\cite{Bossard:2025jjw} and higher-spin UV completions may also be amenable to study with on-shell methods. 

As shown for the massive spin-3/2 examples in this work, on-shell methods avoid the complication of off-shell fields and Lagrangians to reveal the underlying reason for why certain interactions and effective theories are allowed. It is interesting that interactions with gravity are required for massive spin-3/2 particles, implying that broken supergravity is the unique effective theory. Having such a simple and efficient method provides a useful and more transparent way to study gravitational interactions of massive spin-3/2 particles.
 
\vspace{4mm}
\noindent
{\bf Note added}: While this paper was in preparation we became aware of an independent study~\cite{positivity} of massive spin-3/2 amplitudes which uses positivity arguments to also show the necessity of introducing a spin-2 graviton to obtain effective theories with a Planck scale unitarity cutoff.

\section*{Acknowledgements}

We thank Brando Bellazzini, Karim Benakli, Pierre Fayet and Alex Pomarol for helpful discussions. This work is supported in part by the Department of Energy under Grant No.~DE-SC0011842 at the University of Minnesota. The work of T.G.~was partially supported by grant NSF PHY-2309135 to the Kavli Institute for Theoretical Physics (KITP) and the Munich Institute for Astro-, Particle and BioPhysics (MIAPbP) which is funded by the Deutsche Forschungsgemeinschaft (DFG, German Research Foundation) under Germany's Excellence Strategy – EXC-2094 – 390783311.

\appendix

\section{Conventions and useful relations}
\label{app1}

In this appendix we review conventions and useful relations that are used in the calculation of amplitudes.

 \subsection{Spinor-helicity formalism} 
\label{app:spinor}

We assume the metric convention $(+,-,-,-)$, with $\gamma$ matrices defined by
\begin{equation}
    \gamma^\mu= \begin{pmatrix}  0 &  \sigma^\mu_{\alpha\dotalpha} \\
        \bar{\sigma}^{\mu \beta\dotbeta}& 0 
    \end{pmatrix} ,\quad  
    \gamma^5=-i \gamma_0\gamma_1\gamma_2\gamma_3= \begin{pmatrix}  -I_2 & 0 \\
        0 & I_2 
    \end{pmatrix}\,,
 \end{equation}
where $\sigma^\mu=(I_2,\sigma^i)$, $\sigmabar^\mu=(I_2,-\sigma^i)$, $ \sigma^i$ are Pauli matrices and $I_2$ is a $2\times 2$ identity matrix. 

The momentum four-vector $p^\mu $ of a particle can be represented by a bi-spinor $p_{\alpha\dotalpha}\equiv \sigma^\mu_{\al\dotalpha}p_\mu$. In the massless case, we have $p^2=0$ so that the matrix $p_{\alpha\dotalpha}$ has rank $1$, and one can decompose it into chiral ($\lambda_\alpha$)  and anti-chiral ($\tilde{\lambda}_{\dotalpha}$) spinors:\begin{equation}
    p_{\alpha\dotalpha}=\lambda_\al \tilde{\lambda}_{\dotalpha}=\left|p\right>_\alpha\left[p\right|_{\dotalpha},\quad 
    \bar{p}^{\dotalpha\alpha}=\lambda^\al \tilde{\lambda}^{\dotalpha}=\left|p\right]^{\dotalpha}\left<p\right|^\alpha\,.
\end{equation}    
The spinor indices are raised and lowered by the matrix $\epsilon_{\alpha\dotalpha}$, defined by
\begin{equation}
   \epsilon_{\alpha\dotalpha}= \begin{pmatrix}  0 &  -1 \\1& 0 
    \end{pmatrix}\,.
    \label{epmat}
\end{equation}Due to the anticommutation property of the spinors, we have $\left<p\right|^\alpha\left|p\right>_\alpha\equiv \left<pp\right>=0$, and $\left[p\right|_{\dotalpha}\left|p\right]^{\dotalpha}\equiv \left[pp\right]=0$. 
Therefore, the massless spinors satisfy the Dirac equation:  $p\left|p\right>=  \left[p\right|\bar{p}=0$. For simplicity, we drop the spinor indices whenever there is no confusion.  

Remarkably, from the little-group covariance of the amplitudes, the on-shell three-point amplitude is uniquely determined (up to a global factor) by the external particle helicities $h_i$.  
 When $h_1+h_2+h_3>0$, it is given by
\begin{equation}
    M^{h_1 h_2 h_3}=c_+ [12]^{h_1+h_2-h_3}[23]^{h_2+h_3-h_1}[31]^{h_3+h_1-h_2}\,,
    \label{masslesscplus}
\end{equation}
while for $h_1+h_2+h_3<0$ the amplitude is
\begin{equation}
     M^{h_1 h_2 h_3}=c_-\left<12\right>^{h_3-h_1-h_2}\left<23\right>^{h_1-h_2-h_3}\left<31\right>^{h_2-h_3-h_1}\,,
\label{masslesscminus}
\end{equation}
where $c_\pm$ are coupling coefficients.

In the massive case where $p^2=m^2$, the particles transform under the little group $SU(2)$. The momentum bi-spinor can be decomposed as~\cite{Arkani-Hamed:2017jhn}
\begin{equation}   
p_{\alpha\dotalpha}=\epsilon_{IJ}\lambda^I_\alpha\Tilde{\lambda}^J_{ \dotalpha}=\lambda^I_\alpha\Tilde{\lambda}_{I \dotalpha}\equiv \left|\bp\right>^I_\alpha\left[\bp\right|_{I\dotalpha},\qquad     \bar{p}^{\dotalpha\alpha} =\epsilon^{IJ}\Tilde{\lambda}_I^{\dotalpha}\lambda_J^{\alpha}=\Tilde{\lambda}_I^{\dotalpha}\lambda^{I\alpha}\equiv \left|\bp\right]_I^{\dotalpha}\left<\bp\right|^{I \alpha}\,,
\label{mom-decom}
\end{equation}
where the little-group indices $I,J$ are raised and lowered by the $\varepsilon_{IJ}$ tensor given in \eqref{epmat}. 
The spinors satisfy the massive Dirac equations
\begin{equation}
    p|\bp]=m\left| \bp\right>,\quad\bar{p}\left|\bp\right>=m\left| \bp\right],\quad \left[\bp\right|\Bar{p}=-m\left<\bp\right|,\quad \left<\bp\right|p=-m[\bp|\,.
    \label{diraceq-braket}
\end{equation}
In this formalism, particles are conveniently represented by the above bold spinors. For example, the spin-$1/2$ Dirac spinors are
\begin{equation}
  \begin{aligned}
& u^I(p)=\left(\begin{array}{c}
           \lambda^I_\alpha\\
           \Tilde{\lambda}^{I\dotalpha}
    \end{array}\right)=\left(\begin{array}{c}
           \left|\bp\right>\\
           \left|\bp\right]
    \end{array}\right),\quad 
    v^I(p)=\left(\begin{array}{c}
           \lambda^I_\alpha\\
     -      \Tilde{\lambda}^{I\dotalpha}
    \end{array}\right)=\left(\begin{array}{c}\left|\bp\right>\\
    -       \left|\bp\right]
    
    \end{array}\right)\,,\\
    &\bar{u}_I(p)=\left(-\lambda_I^{\alpha}\quad \Tilde{\lambda}_{I\dotalpha}\right)=\left(-\left<\bp\right|\quad \left[\bp\right|\right),\quad \bar{v}_I(p)=\left(\lambda_I^{\alpha}\quad \Tilde{\lambda}_{I\dotalpha}\right)=\left(\left<\bp\right|\quad \left[\bp\right|\right)\,.
  \end{aligned} \label{uvlam}
\end{equation} The polarization vector for a massive spin-1 particle of momentum $p$  is given by Eq.~\eqref{eq:massivepolvect}. Higher-spin  polarization vectors are simply products of $\varepsilon$'s and $u$ (see, for example, \cite{Chung:2018kqs}).   In particular, spin-$3/2$ is a vector-spinor constructed from a spin-1 polarization vector and a spin-$1/2$ spinor
\begin{equation}
    \bar{v}_\mu^{IJK}(\bp)=\left(\begin{array}{cc}
        \left<\bp\right| \bar{\varepsilon}_\mu& \left[\bp\right| \bar{\varepsilon}_\mu 
    \end{array}\right),\quad   u_\mu^{IJK}(\bp)=\left(\begin{array}{c}
         \left|\bp\right> \varepsilon_\mu \\
\left|\bp\right]\varepsilon_\mu
    \end{array}\right)\,,
    \label{eq:polvec32}
\end{equation}
where $IJK$ are symmetrized indices. This construction automatically ensures the Dirac equation, the trace constraint $\gamma^\mu u_\mu=\bar{v}_\mu\gamma^\mu =0$ and the divergence constraint $p^\mu u_\mu=\bar{v}_\mu p ^\mu =0$.

It is convenient to choose a special $SU(2)$ basis, where we write the massive spinors $\lambda_\alpha^I$, $\tilde{\lambda}_{\dotalpha}^I$ in terms of $\lambda_\al, \Tilde{\eta}_{\dotalpha}$   (of $-1/2$ helicity), and $\Tilde{\lambda}_{\dotalpha}, \eta_\al$ (of $+1/2$ helicity)
\begin{equation}   
\begin{aligned}        
&\lambda^I_\alpha=\lambda_\alpha\zeta^{-I}+\eta_\alpha\zeta^{+I }\,, \\&\Tilde{\lambda}^I_{\dotalpha}=\Tilde{\lambda}_{\dot\alpha}\zeta^{+I}+\Tilde{\eta}_{\dotalpha}\zeta^{-I }\,.
\end{aligned}
\label{decomp-lam}
\end{equation}
They satisfy
\begin{equation}   
\zeta^{-I}\zeta^+_I=1,\quad  \lambda^\alpha\eta_\alpha = \Tilde{\lambda}_{\dotalpha}\Tilde{\eta}^{\dotalpha}=m\,.
\end{equation}
In this basis, the momentum is
\begin{equation} p_{\alpha\dotalpha}=\lambda_\al\Tilde{\lambda}_{\dotalpha}-\eta_\al\Tilde{\eta}_{\dotalpha}\,.
    \label{momentumlam}
\end{equation} 
The spinors $\lambda_\al, \Tilde{\eta}_{\dotalpha}$, $\Tilde{\lambda}_{\dotalpha}, \eta_\al$ are also used to represent polarization vectors of different helicities. For spin-1, the polarization vectors are given in  \eqref{eq:masslesspolvect}. Similarly for spin-$3/2$, the helicities $\pm3/2$, $\pm1/2$ are decomposed in this basis with appropriate Clebsch-Gordan coefficients as
\begin{equation}
    \begin{aligned}
     &   u^{+\frac{3}{2}}=\frac{\sqrt{2}}{m_{3/2}}\left(\begin{array}{c}\Tilde{\lambda}_{\dotalpha}\eta_\alpha\eta_\beta  \\
    \Tilde{\lambda}_{\dotalpha} \eta_\alpha\Tilde{\lambda}^{\dotbeta}
        \end{array}\right),\quad u^{+\frac{1}{2}}=\frac{\sqrt{2/3}}{m_{3/2}}\left(\begin{array}{c}\lambda_\alpha            \Tilde{\lambda}_{\dotalpha}\eta_\beta 
 +\lambda_\beta               \Tilde{\lambda}_{\dotalpha}\eta_\alpha +\eta_\alpha\eta_\beta\tilde{\eta}_{\dotalpha}
 \\\eta_\alpha               \Tilde{\eta}_{\dotalpha}\Tilde{\lambda}^{\dotbeta}+\eta_\alpha               \Tilde{\eta}^{\dotbeta} \Tilde{\lambda}_{\dotalpha}+\tilde{\lambda}_{\dotalpha}\tilde{\lambda}^{\dotbeta}\lambda_\alpha
        \end{array}\right)\,,\\&  u^{-\frac{3}{2}}=\frac{\sqrt{2}}{m_{3/2}}\left(\begin{array}{c}
               \tilde{\eta}_{\dotalpha}\lambda_\alpha\lambda_\beta  \\               \tilde{\eta}_{\dotalpha}\lambda_\alpha\Tilde{\eta}^{\dotbeta}
        \end{array}\right),\quad      u^{-\frac{1}{2}}=\frac{\sqrt{2/3}}{m_{3/2}}\left(\begin{array}{c}\lambda_\alpha               \Tilde{\lambda}_{\dotalpha}\lambda_\beta 
 +\eta_\beta                {\lambda}_{\alpha}\tilde{\eta}_{\dotalpha} +\lambda_\alpha\eta_\beta\tilde{\eta}_{\dotalpha}
 \\\lambda_\alpha               \Tilde{\lambda}_{\dotalpha}\Tilde{\eta}^{\dotbeta}+\lambda_\alpha               \Tilde{\lambda}^{\dotbeta} \Tilde{\eta}_{\dotalpha}+\tilde{\eta}_{\dotalpha}\tilde{\eta}^{\dotbeta}\eta_\alpha
        \end{array}\right)\,.
    \end{aligned}
\end{equation}

The massive amplitudes also transform under the $SU(2)$ little-group, and similar to the massless case \eqref{masslesscplus}-\eqref{masslesscminus}, there exists a systematic classification of massive three-point amplitudes in terms of the mass and the spin of the external legs \cite{Arkani-Hamed:2017jhn}.  For an amplitude containing a massive spin-$s$ particle, one can strip off all the little-group indices to obtain 
\begin{equation}
    M^{I_1\cdots I_{2s}}=\lambda_{\alpha_1}^{I_1}\cdots \lambda_{\alpha_{2s}}^{I_{2s}}M^{\alpha_1\cdots \alpha_{2s}}\,,
\end{equation}
where $\{\alpha_1,\cdots ,\alpha_{2s}\}$ are totally symmetric and $\{I_1,\cdots, I_{2s}\}$ are the little-group indices for the spin-$s$ particle. The  classification is based on the fact that the stripped amplitude only carries $SL(2,\mathbb{C})$ indices, and thus the question of finding a general amplitude reduces to choosing an appropriate basis $(v_\alpha,u_\alpha)$ that spans  the $SL(2,\mathbb{C})$ space. The choice of basis depends on the mass of the three external legs.

Finally, in spinor manipulations, we have applied several useful relations. 
Two-component spinors  satisfy the Schouten identities
\begin{equation}
\begin{aligned}&
|\bthree][\bone\btwo] +|\bone][\btwo\bthree]+|\btwo][\bthree\bone]=0\,,\\&
\left|\bthree\right>\left<\bone\btwo\right> +\left|\bone\right>\left<\btwo\bthree\right>+\left|\btwo\right>\left<\bthree\bone\right>=0\,.
\end{aligned}
\label{schouten}
\end{equation}
For the three-point amplitude involving one massless particle and two massive particles of equal mass, one  introduces the so-called   $x$-factor \cite{Arkani-Hamed:2017jhn}, defined in \eqref{xfactordef}. Using this definition, along with the Schouten identities, momentum conservation, and the Dirac equation, one can derive the following useful relations
\begin{equation}
  \begin{aligned}
  & x\left<\btwo\bthree\right>=\left<\btwo \varepsilon_1^+\bthree\right]+\left<\bthree \varepsilon_1^+\btwo\right],\quad \frac{1}{x}\left<\btwo\bthree\right>=\left<\btwo \varepsilon_1^-\bthree\right]+\left<\bthree \varepsilon_1^-\btwo\right]+\frac{1}{m}\left<1\btwo\right>\left<1\bthree\right>\,,
  \\& x\left[\btwo\bthree\right]=\left<\btwo \varepsilon_1^+\bthree\right]+\left<\bthree \varepsilon_1^+\btwo\right]+\frac{1}{m}[1\btwo][1\bthree]
 ,\quad \frac{1}{x}\left[\btwo\bthree\right]=\left<\btwo \varepsilon_1^-\bthree\right]+\left<\bthree \varepsilon_1^-\btwo\right]\,, 
 \\&x\left<1\btwo\right>=-\left[1\btwo\right],\quad \frac{1}{x} \left[1\btwo\right]=-\left<1\btwo\right>\,.
  \end{aligned}
  \label{xrelation}  
\end{equation}
 
\subsection{Explicit kinematics}
\label{app:explicitkin}

Our amplitudes are evaluated with explicit kinematics in the center-of-mass frame. The four-momentum is defined by $p^\mu\equiv (E,\Vec{p})$, where the spatial component is projected onto spherical coordinates  as
\begin{equation}
    \Vec{p}=\left(p \sin \theta\cos\phi,p\sin\theta\sin\phi,p\cos\theta\right)\,,
\end{equation}
where $p=|
\vec{p}|$, $0\leq \phi \leq 2\pi$ and $0\leq \theta \leq \pi$. In the basis \eqref{momentumlam}, the momentum bi-spinor is a combination of two-component spinors $\lambda,\tilde{\lambda},\eta,\tilde{\eta}$, which are expressed in spherical coordinates as
\begin{equation}
    \begin{aligned}
  &\lambda_\alpha=\sqrt{E+p}\left(\begin{array}{c}
             -s^* \\c
     \end{array}\right)\,,\qquad \Tilde{\lambda}_{\dotalpha}=\sqrt{E+p}\left(\begin{array}{c}
       -s \\c^*
     \end{array}\right)\,,\\&\eta_\alpha=\sqrt{E-p}\left(\begin{array}{c}
             c^* \\s
        \end{array}\right)\,,\qquad \Tilde{\eta}_{\dotalpha}=-\sqrt{E-p}\left(\begin{array}{c}
             c\\s^*
        \end{array}\right)\,,
    \end{aligned} 
\end{equation}with
\begin{equation}
    s\equiv \sin \frac{\theta}{2}\,e^{i\frac{\phi}{2} },\quad c\equiv\cos \frac{\theta}{2}\,e^{i\frac{\phi}{2}}\,.
\end{equation}
The polarization vectors are 
\begin{equation}
    \begin{aligned}
        &\varepsilon^{\mu+}=-\frac{1}{\sqrt{2}}(0,\cos \theta\cos \phi-i \sin \phi,\cos \theta\sin \phi+i \cos \phi,-\sin \theta)\,,\\  &\varepsilon^{\mu-}=\frac{1}{\sqrt{2}}(0,\cos \theta\cos \phi+i \sin \phi,\cos \theta\sin \phi-i \cos \phi,-\sin \theta)\,,\\&\varepsilon^{\mu 0}=\frac{1}{m}(p,E\sin \theta \cos \phi,E\sin \theta \sin \phi, E \cos \theta)\,.
    \end{aligned}\label{explicitep}
\end{equation}
They are normalized by $\varepsilon^+\cdot \varepsilon^-=1$, $\varepsilon^0\cdot \varepsilon^0=-1$. The Dirac spinors defined in \eqref{diraceq-braket} become
\begin{equation}
\begin{aligned}
    u^+ = \begin{pmatrix}
        \sqrt{E-p}c^*\\  \sqrt{E-p}s\\  \sqrt{E+p}c^*\\\sqrt{E+p}s
    \end{pmatrix}  ,\quad u^- = \begin{pmatrix}
        -\sqrt{E+p}s^*\\  \sqrt{E+p}c\\ - \sqrt{E-p}s^*\\\sqrt{E-p}c
    \end{pmatrix}\,. 
\end{aligned}
\end{equation}

We assume all momenta to be incoming $\sum_i p_i=0$, so the final states have negative energy. The corresponding polarization vectors for the final states are then obtained by analytical continuation: $\left|-\bp\right>=-\left|\bp\right>$, $\left|-\bp\right]=\left|\bp\right]$. In terms of explicit kinematics, this can be realized by
\begin{equation}
    p^0\rightarrow -p^0 ,\quad p\rightarrow -p ,\quad \theta \rightarrow \theta,\quad \phi\rightarrow \phi\,.
\end{equation}   
In the center-of-mass frame, the angles are given by
\begin{equation}\begin{aligned}
    &\theta_1=\pi,\quad \phi_1=0, \quad \theta_2=0,\quad \phi_2=0, \\&\theta_3=\theta-\pi,\quad \phi_3=0, \quad \theta_4=\theta,\quad \phi_4=0\,.
    \end{aligned}
    \label{comangles}
\end{equation}

\section{All-line-transverse momentum shift}
\label{app:shift}

In this appendix we review the all-line-transverse momentum shift~\cite{Ema:2024vww} that is used to construct the four-point amplitudes.
The three-point amplitudes in the massless and massive cases are dictated by the little-group covariance, and are independent of the Lagrangian. 
The construction of higher-point amplitudes from three-point building blocks is realized by the recursion relation. Momentum is shifted by a complex variable $z$: $p_i\rightarrow \hat{p}_i= p_i+ z r_i$, and consequently, the $n$-point amplitude complexifies ${\cal A}_n\rightarrow \widehat{\cal A}_n(z)$. When $z=0$, we recover the original amplitude.

Using the Cauchy theorem, the contour integral of $\widehat{\cal A}_n(z)/z$ is given by the residue of this function at the poles. For a small contour around $z=0$, one obtains the amplitude ${\cal A}_n=\widehat{\cal A}_n(0)$, and when the contour is enlarged to $z\rightarrow\infty$, the integral picks up all the poles as well as a \textit{boundary term} $B_\infty$, if the function $\widehat{\cal A}_n(z)/z$ does not vanish at complex infinity.  We have
\begin{equation}
\begin{aligned}
	 {\cal A}_n &= \frac{1}{2\pi i}\oint_{z=0} \frac{dz}{z} \widehat{\cal A}_n(z)
  = - \sum_{\{z_I\}} \text{Res}\left[\dfrac{\widehat{\cal A}_n(z)}{z}\right] + B_\infty\,.
  \label{eq:CauchyAn}
\end{aligned}
\end{equation}
By locality, the complex amplitude has physical poles  $z_I$  when the intermediate particle is on-shell. Furthermore, the pole structure indicates that at the physical poles,  the amplitude factorizes into subamplitudes \cite{Weinberg:1995mt}
\begin{equation}
    \begin{aligned}
	{\cal A}_n = -\sum_{z = z_I}\sum_\lambda\mathrm{Res}\left[\widehat{\cal A}^{(\lambda)}_{n-m+2}\frac{1}{z}\frac{1}{\hat{p}_I^2 - m_I^2}
	\widehat{\cal A}^{(-\lambda)}_{m}\right]
	+ B_\infty\,,
\end{aligned}
\label{npointconstruction}
\end{equation}
where $m<n$ and $\lambda$, $p_I$, $m_I$ are the helicity, momentum and mass of the intermediate particle, respectively. The values $z_I$ are solutions of the intermediate particle on-shell relation: $\hat{p}^2(z_I)=m_I^2$. In particular, if $B_\infty=0$, the four-point amplitude becomes
\begin{equation}
\begin{aligned}
	{\cal A}_4 = -\sum_{z = z_I}\sum_\lambda\mathrm{Res}\left[\widehat{\cal A}^{(\lambda)}_{3}\frac{1}{z}\frac{1}{\hat{p}_I^2 - m_I^2}
	\widehat{\cal A}^{(-\lambda)}_{3}\right]\,.
\end{aligned} 
\label{four-point-fact}
\end{equation}
The above expression demonstrates the essential idea of recursion: from on-shell three-point amplitudes that are determined by  the little group, one can construct four- and higher-point amplitudes using the residue formula. The factorization does not depend on any assumption about the Lagrangian. However, for the amplitude calculation to be unambiguous, it is crucial that the boundary term $B_\infty$ vanishes, and given that the complex structure arises from the initial momentum shift, the latter will determine whether or not a theory is \textit{constructible}.

In the massless case, the Britto-Cachazo-Feng-Witten (BCFW) momentum shift~\cite{Britto:2004ap,Britto:2005fq} allows for a recursive construction of $n$-point graviton and gluon amplitudes. For massive amplitudes, an all-line-transverse (ALT) shift was recently proposed in \cite{Ema:2024vww,Ema:2024rss}, which induces no boundary term in QED, the electroweak theory and supergravity \cite{Gherghetta:2024tob}.
The ALT shift is defined by  $p_i\rightarrow \hat{p}_i= p_i+ z r_i$ with $r_i={\bar c}_i\varepsilon_i^+ ({\bar c}_i\varepsilon_i^-)$ for an external particle $i$ with positive (negative) helicity, where ${\bar c}_i$ are constants, and  $\varepsilon^\pm_i$ are the transverse polarization vectors for the particle $i$. This shift applies to all particles.  

In the basis  \eqref{decomp-lam},  the transverse polarization vectors are given in \eqref{eq:masslesspolvect}, thus the ALT shift can be equivalently defined as
\begin{equation}
    \lambda_{i\alpha}\rightarrow   \lambda_{i\alpha}+ w \eta_{i\alpha}\,,
    \quad \Tilde{\eta}_{i,\dotalpha}\rightarrow \Tilde{\eta}_{i,\dotalpha}+\tilde{w} \Tilde{\lambda}_{i,\dotalpha}\,,
    \label{positiveshift}
\end{equation}
if particle $i$ has positive helicity, and
\begin{equation}
\eta_{i\alpha}\rightarrow   \eta_{i\alpha}+\tilde{w}  \lambda_{i\alpha}\,,
\quad \Tilde{\lambda}_{i,\dotalpha}\rightarrow \Tilde{\lambda}_{i,\dotalpha}+w \Tilde{\eta}_{i,\dotalpha}\,,
\label{negativeshift}
\end{equation}
 if particle $i$ has negative helicity.  The constants satisfy the relation $w-\tilde{w}=\sqrt{2}z{\bar c}_i/m$. Next, we state several important properties of the ALT shift.  As mentioned in Appendix \ref{app1}, since $\tilde{\lambda},\eta$ ($\tilde{\eta},\lambda$) have helicity $+1/2$ ($-1/2$), the transverse polarization vector of helicity $+s$ ($-s$) for a particle of spin $s$ is given by a combination of $2s$ $\tilde{\lambda},\eta$ ($\tilde{\eta},\lambda$). Then by construction, the ALT shift only changes the spinors that are not in the combination, and therefore the transverse polarization vector is not shifted.\footnote{In fact, the longitudinal polarization vector for spin-1 is not shifted,   if one chooses specific $w,\tilde{w}$ \cite{Ema:2024rss}. However, the longitudinal polarization for particles with spin $>1$ is shifted \cite{Gherghetta:2024tob}.}

Furthermore, due to  transversality $p_i\cdot \varepsilon_i^\pm=0$ and $\varepsilon_i^\pm\cdot\varepsilon_i^\pm=0$, the momentum after the shift still satisfies the on-shell condition $\hat{p}_i\cdot\hat{p}_i=p_i\cdot p_i=m_i^2$. Besides, the total momenta after the shift are conserved
\begin{equation}
    \sum_i \hat{p}_i=    \sum_i  {p}_i +z\sum_ir_i =0\quad \leftrightarrow \quad \sum_ir_i =0\,.
    \label{consshift}
\end{equation}
This is because $r_i={\bar c}_i\varepsilon_i^\pm$ and in terms of explicit kinematics \eqref{explicitep},  the time-component of the transverse polarization vector is zero.  Eq.~\eqref{consshift}  then corresponds to three equations for at least four parameters $\{{\bar c}_i\}$, which always has a solution. 

To show that a theory is constructible under the ALT shift, we use dimensional analysis. Generically, a tree-level $n$-point amplitude can be decomposed into\begin{equation}
    {\cal A}_n = \left(\sum_{\mathrm{diagrams}}g \times F\right) \times \prod_{\mathrm{vectors}} \varepsilon \times \prod_{\mathrm{fermions}}u\,,
    \label{eq:An}
\end{equation}
where $g$ represents the coupling coefficients, $F$ encodes the kinematic factors, including the propagator and momentum insertions, and  $\varepsilon, u$ are the boson polarization vector and fermion wavefunction, respectively. A spin-$3/2$ particle is composed of the product $\varepsilon \times  u$. We will assume that all external states are \textit{transverse}, and after constructing the amplitude, the longitudinal states can be obtained by applying spin raising or lowering operators. As we have argued, $\prod  \varepsilon \times \prod u$ are not shifted, so all the $z$ dependence arises from $F$.

Using dimensional analysis on \eqref{eq:An} implies
\begin{equation}
    4-n = [g] + [N] - [D] + \frac{1}{2}N_f\,,
\end{equation}
where $N_f$ is the number of fermions and $F\equiv N/D$. Since all the momenta are linearly shifted, the propagator $D$ behaves as $1/D\sim z^{-[D]}$ at large $z$. The numerator $N$ contains masses and momentum insertions, therefore, $N\sim z^{\gamma_N}$ with $\gamma_N\leq [N]$. Finally, the large-$z$ behavior of the $n$-point amplitude is $\widehat{\cal A}_n\sim z^\gamma$ with
\begin{equation} 
	\gamma = \gamma_N - [D] \leq [N] - [D] = 4-n - [g] - \frac{1}{2}N_f\,.
 \label{naivegamma}
 \end{equation}
For instance, consider the four-point amplitude of $e^+e^-$ scattering in QED, where $n=4$, $[g]=0$ and $N_f=4$ and therefore $\gamma\leq -2$ meaning that $\widehat{\cal A}(z)\rightarrow0 $ at complex infinity. There is no boundary term and the amplitude is constructible. For four-point gravitino scattering in the presence of gravity, the coupling is dimensionful $[g]=[\kappa^2]=-2$ and $N_f=4$, $n=4$, so naively, $\gamma\leq0$. The boundary term seems to be present when $\gamma=0$.

However, if the four-point amplitude satisfies the Ward identity in the massless limit, the actual large-$z$ behavior is improved. The highest $z^0$ order is attained when all the momenta are replaced by the shifted ones $r_i$, which are massless. This order then corresponds to the amplitude with massless kinematics. Moreover, since $r_i\propto \varepsilon_i^\pm$, the external polarization can be replaced by the momentum $r_i$, and imposing the Ward identity, the massless amplitude then vanishes. We are left with the next order $z^{-1}$, so the gravitino scattering amplitude is also constructible.

\section{Spin-3/2 contact terms from photon exchange}
\label{app:contact}

In this appendix we provide further details on recovering the spin-3/2 contact terms from photon exchange in Section~\ref{sec41}.
To search for the contact terms giving rise to the amplitude \eqref{firstfrr}, we can either start from \eqref{firstfrr} and use relations of $r_i$ to recover $\varepsilon_i$, or write down a general set of contact terms and replace $\varepsilon_i$ by $c_ir_i$, to match  \eqref{firstfrr}. 

We adopt here the first approach. Since the shifted momenta $r_i$ are massless and satisfy momentum conservation, we have the relation
\begin{equation}
\begin{aligned}
        (r_1\cdot r_2)(r_1-r_2)\cdot r_3
        =&\,(r_1\cdot r_2) (2r_1\cdot r_3+r_3\cdot r_4)\,,\\
        =&\,(r_1\cdot r_2)(r_3\cdot r_4)-2(r_1\cdot r_3 ) (r_1\cdot r_3)-2(r_1\cdot r_3 ) (r_1\cdot r_4)\,,\\
        =&\,(r_1\cdot r_4)(r_2\cdot r_3)-(r_1\cdot r_3)(r_2\cdot r_4)\,.
\end{aligned}
\label{rrelation1}
\end{equation}
In the last line,  we used the identities $r_1\cdot r_3=r_2\cdot r_4$, and \begin{equation}
    2 (r_1\cdot r_3)(r_1\cdot r_4)=(r_1\cdot r_2)(r_3\cdot r_4)-(r_1\cdot r_4)(r_2\cdot r_3)-(r_1\cdot r_3)(r_2\cdot r_4)\,.
    \label{rrelation2}
\end{equation}
Using \eqref{rrelation1}, \eqref{firstfrr} can then be written in the form
\begin{equation}
-    \frac{f_2}{2 r_1\cdot r_2}= \frac{1}{M^2} (\bar{v}_2 u_1)(\bar{v}_3 u_4)(l_1+2l_2)^2 \left[(\varepsilon_1\cdot \varepsilon_4)(\varepsilon_2\cdot \varepsilon_3)-(\varepsilon_1\cdot \varepsilon_3)(\varepsilon_2\cdot \varepsilon_4)\right]\,.
\label{frrsimplify}
\end{equation}
This corresponds to the contact term $\xi_1$ in \eqref{contactg1}, when $\xi_2=0$. The expression \eqref{firstfrr} can be further simplified by using the following relation
\begin{equation}
    \begin{aligned}       &\left(\bar{v}_2u_1\right)\left(\bar{v}_3u_4\right)\left(r_1\cdot r_2\right)\left(r_1-r_2\right)\cdot r_3  \\
    =&\,\frac{1}{2}       \left(\bar{v}_2u_1\right)\left(\bar{v}_3u_4\right)\left(r_3\cdot r_4\right)\left(r_1\cdot r_3-r_1\cdot r_4\right)+\frac{1}{2}      \left(\bar{v}_2u_1\right)\left(\bar{v}_3u_4\right)\left(r_1\cdot r_2\right)\left(r_1\cdot r_3-r_1\cdot r_4\right)\,,\\
    =&\,\frac{1}{4}  \left(\bar{v}_2u_1\right)\left(r_3\cdot r_4\right)\bar{v}_3\gamma_\mu\gamma_\nu \left( {r}_2^\mu {r}_1^\nu- {r}_1^\mu {r}_2^\nu\right)u_4+\frac{1}{4}\left(\bar{v}_3u_4\right)\left(r_1\cdot r_2\right)\bar{v}_2\gamma_\mu\gamma_\nu\left( {r}_3^\mu {r}_4^\nu- {r}_4^\mu {r}_3^\nu\right)u_1\,,\\
    =&\,\frac{1}{2} \left(\bar{v}_2u_1\right)\left(r_3\cdot r_4\right)\bar{v}_3\gamma_{\mu\nu}{r}_2^\mu {r}_1^\nu u_4+\frac{1}{2}\left(\bar{v}_3u_4\right)\left(r_1\cdot r_2\right)\bar{v}_2\gamma_{\mu\nu}   {r}_3^\mu {r}_4^\nu u_1\,.
    \end{aligned}
    \label{eq:r123rel}
\end{equation}
In the second equality, we used the anticommutation relation $\left\{\gamma^\mu,\gamma^\nu\right\}=2\eta^{\mu\nu}$ and the equations of motion $\bar{v}_i \slashed{r}_i=\slashed{r}_i u_i=0$. Substituting \eqref{eq:r123rel} into \eqref{firstfrr} gives the expression
\begin{equation}
-\frac{f_2}{2 r_1\cdot r_2}= \frac{1}{2M^2}  (l_1+2l_2)^2  \left[ \left(\bar{v}_2^\mu u_1^\nu\right) \bar{v}_3^\al\gamma_{\mu\nu}      u_{4\al} + \left(\bar{v}^\mu_3u_4^\nu\right) \bar{v}_2^\al\gamma_{\mu\nu}u_{1\al}\right]\,.
 \end{equation}
This corresponds to the $\xi_2$ contact term \eqref{contactg2}, if $\xi_1=0$. In the general case, for $\xi_1,\xi_2\neq0$, to reproduce the amplitude \eqref{firstfrr} from the sum of  both contact terms, the coupling should satisfy
\begin{equation}
    \xi_1+2\xi_2=\frac{1}{M^2}(l_1+2l_2)^2\,.
\end{equation}

\section{Minimal $N=2$ supergravity Lagrangian}
\label{app:sugra}

In this appendix we write down the minimal $N=2$ Lagrangian for comparison with our bottom-up results. The $N=2$ supergravity Lagrangian was first constructed by coupling a massless $\left(\frac{3}{2},1\right)$ multiplet to the gravity multiplet $\left(2,\frac{3}{2}\right)$ \cite{Ferrara:1976fu}. The minimal Lagrangian $\mathcal{L}_{N=2}=\mathcal{L}_G+\mathcal{L}_M+\mathcal{L}_{4f}$ is composed of an ordinary $N=1$ supergravity piece $\mathcal{L}_G$, a matter action $\mathcal{L}_M$ and additional four-fermion contact terms $\mathcal{L}_{4f}$, respectively given by
\begin{equation}
    \begin{aligned} 
 &  \mathcal{L}_G=\frac{2}{\kappa^2}VR(V)+\frac{i}{2}V\bar{\psi}_\mu \gamma^{\mu\nu\rho} D_\nu\psi_\rho  
  \\& \mathcal{L}_M=\frac{i}{2}V\bar{\chi}_\mu \gamma^{\mu\nu\rho} D_\nu\chi_\rho -\frac{1}{4}V g^{\mu\rho}g^{\nu\sigma}F_{\mu\nu}F_{\rho\sigma}+\frac{\kappa}{\sqrt{2}}\psib_\mu (VF^{\mu\nu} -i\gamma^5 \widetilde{F}^{\mu\nu})\psi_\nu\,,
   \\
   & \mathcal{L}_{4f}=-\frac{1}{4}\kappa^2\bar{\psi}_\mu \chi_\nu \left[ V(\bar{\psi}^\mu\chi^\nu -\bar{\psi}^\nu\chi^\mu ) +\epsilon^{\mu\nu\alpha\beta}(\bar{\psi}_\alpha \gamma^5 \chi_\beta)\right]\\
   &\qquad -\frac{V}{16}\kappa^2\left[(\bar{\psi}^\mu\gamma^\nu\psi^\alpha)(\bar{\chi}_\mu\gamma_\nu\chi_\alpha+2\bar{\chi}_\nu\gamma_\mu\chi_\alpha)-4(\bar{\chi}_\mu\gamma_\nu\chi^\nu) (\bar{\psi}^\mu\gamma_\alpha\psi^\alpha) \right] \\
   &\qquad-\frac{V}{32}\kappa^2\left[(\bar{\chi}^\mu\gamma^\nu\chi^\alpha)(\bar{\chi}_\mu\gamma_\nu\chi_\alpha+2\bar{\chi}_\nu\gamma_\mu\chi_\alpha)-4(\bar{\chi}_\mu\gamma_\nu\chi^\nu) ^2\right] \\&\qquad -\frac{V}{32}\kappa^2\left[(\bar{\psi}^\mu\gamma^\nu\psi^\alpha)(\bar{\psi}_\mu\gamma_\nu\psi_\alpha+2\bar{\psi}_\nu\gamma_\mu\psi_\alpha)-4(\bar{\psi}_\mu\gamma_\nu\psi^\nu) ^2\right]\,,
   \end{aligned}
\label{n2lagrangian}
\end{equation}
where $V$ is the determinant of the vielbein, i.e. $\det (V_\mu^\alpha)=V$, $D_\mu$ is the torsionless gravitational covariant derivative, and the dual field strength $\widetilde{F}_{\mu\nu}\equiv\frac{1}{2}\epsilon_{\mu\nu\alpha\beta}F^{\alpha\beta}$.
The supersymmetry transformations for the component fields are given in \cite{Ferrara:1976fu}.

The Lagrangian, $\mathcal{L}_{N=2}$ has a global $U(1)$ symmetry related to the $\chi\rightarrow e^{i \alpha}\chi,\psi\rightarrow e^{i \alpha}\psi$ rotations, and only has a gravitational coupling ($\kappa\equiv 2/M_P$). The global  $U(1)$ symmetry can further be gauged by the massless spin-$1$ field \cite{Freedman:1976aw}, related to the gauge coupling $e$. In this case, local supersymmetry requires that the gauge coupling $e$ is related to the gravitational coupling through $m_{3/2}=\sqrt{2}e M_P\propto\kappa \sqrt{ \Lambda_c}$, where $m_{3/2}$ is a gravitino mass term and $\Lambda_c$ is the cosmological constant.
When supersymmetry is unbroken, despite the apparent mass term, the gravitino is massless. This can be verified by carefully examining the equation of motion in curved spacetime \cite{Deser:1977uq}. However, supersymmetry can be softly broken by turning off the cosmological constant, while keeping the rest of the Lagrangian unchanged, and this corresponds to a theory in Minkowski spacetime with a \textit{massive} charged Dirac gravitino.   It has been shown \cite{Deser:2001dt} that the gravitino in such a theory propagates causally.

The graviton-gravitino coupling can be recovered by expanding the metric around Minkowski space, $g_{\mu\nu}\simeq \eta_{\mu\nu}+\kappa h_{\mu\nu} $, which gives rise to the three-point coupling $\frac{\kappa}{2}h^{\mu\nu}T_{\mu\nu}$ where $h_{\mu\nu}$ is the graviton and $T_{\mu\nu} $ is the energy momentum tensor of the gravitino.  

\bibliographystyle{utphys}
\bibliography{refs}
\end{document}